\documentclass[preprint,11pt]{elsarticle}
\usepackage[margin=1in]{geometry}
\usepackage{amsmath,amssymb,amsfonts,amsthm}
\usepackage{graphicx}
\usepackage{array}
\usepackage{natbib}
\usepackage{lineno}
\usepackage{algorithm}
\usepackage{algpseudocode}
\algrenewcommand{\algorithmicrequire}{\textbf{Input:}}
\algrenewcommand{\algorithmicensure}{\textbf{Output:}}

\newtheorem{proposition}{Proposition}
\newtheorem{corollary}{Corollary}
\newtheorem{definition}{Definition}
\newtheorem{example}{Example}
\newcommand{\E}{\mathbb{E}}
\newcommand{\Prb}{\mathbb{P}}
\newcommand{\ind}{\mathbf{1}}
\date{}
\usepackage{microtype}
\usepackage{placeins}
\usepackage[hidelinks]{hyperref}
\hypersetup{pdftitle={Information Design for Behaviorally Stable Platooning},pdfauthor={Jia Li and Di Chen}}
\begin{document}
\begin{frontmatter}
\title{Information Design for Behaviorally Stable Platooning}
\author[address1]{Jia Li\corref{mycorrespondingauthor}}
\cortext[mycorrespondingauthor]{Corresponding author}
\ead{jia.li1@wsu.edu}
\author[address2]{Di Chen}
\address[address1]{Department of Civil and Environmental Engineering, Washington State University, USA}
\address[address2]{Department of Civil and Environmental Engineering, University of California Davis, USA}

\begin{abstract}
Voluntary platoon formation requires vehicles to coordinate despite uncertain downstream fuel savings and private joining incentives. We study whether an informed operator can encourage participation by designing a public disclosure rule. A finite-vehicle coordination game links physical fuel savings to vehicles' joining decisions under two information environments: private joining costs and private observations of the common corridor state. Private costs yield a posterior-mean participation response, while private observations require richer posterior information and create correlated formation outcomes. We formulate persuasive disclosure problems and develop computational methods for both regimes. In a literature-anchored reference scenario, optimized two-message disclosure supports participation of 8.74\% and 51.83\%, respectively, compared with zero without disclosure and 5.15\% and 39.27\% under full disclosure. Gains concentrate in intermediate opportunities, depend on joining-cost calibration, and need not maximize welfare. Numerical sensitivity and trajectory checks quantify these effects. All gains are conditional on favorable symmetric equilibrium selection: public information alone cannot eliminate the no-participation equilibrium when joining requires a costly commitment without a guaranteed partner.
\end{abstract}
\end{frontmatter}

\FloatBarrier
\section{Introduction}\label{sec:introduction}
Vehicle platooning promises fuel savings by allowing connected and automated vehicles to travel at short following distances. Realizing these savings requires both a technically feasible formation and the voluntary participation of its members. Existing research has developed methods for coordinating compatible vehicle routes and examined the safety requirements of platooning \citep{larsson2015vehicle,axelsson2016safety}. For a self-interested carrier, however, a feasible platoon is only a prospective opportunity. Reaching a rendezvous, adjusting a departure time, or waiting for other vehicles incurs costs, while the resulting savings depend on how much downstream travel supports close following and whether other eligible vehicles also join. A vehicle may therefore decline an opportunity that would be attractive if enough others participated. Understanding this interdependence is necessary for turning potential fuel savings into voluntary platoon formation.

The literature on behaviorally stable platooning provides an important starting point. Unequal fuel savings across platoon positions can undermine vehicles' willingness to form or maintain a platoon. \citet{sun2019behaviorally} address this problem through benefit redistribution, and \citet{sun2021decentralized} develop decentralized game-theoretic coordination mechanisms. Related work also addresses private preferences and strategic behavior: the leader auction of \citet{sun2021auction} accounts for privately known willingness to pay, while \citet{barua2023participation} study participation-maximizing matching with truck preferences and strategic reporting. Thus, self-interest and private information are already recognized in platooning research. These studies establish how the allocation of benefits, positions, or partners can support cooperation; they also motivate a complementary question about the information on which vehicles base their initial participation decisions.

In particular, an operator may know more than an individual vehicle about the common downstream opportunity. Corridor-wide observations can indicate how much of a shared route will permit sustained platooning, whereas a vehicle's local observations provide only partial evidence. Vehicles may additionally know their own scheduling or access costs, which the operator does not observe. Before any platoon forms, the operator must decide what to disclose about the opportunity. A public message changes both a vehicle's expected savings and its expectations of other vehicles' participation. The research question is therefore: \emph{how should an informed operator disclose a common platooning opportunity to encourage collective participation by self-interested vehicles with private costs or private observations, while keeping their participation decisions individually optimal?} The gap addressed here concerns the joint determination of public disclosure and voluntary participation under common uncertainty. The redistribution, auction, and matching mechanisms above do not provide this disclosure characterization, although they could complement it in a broader formation protocol.

We formulate this problem for a finite cohort of $N$ eligible vehicles. The physical payoff values distance-dependent aerodynamic fuel savings against monetary joining costs and explicitly assigns zero platooning benefit to a vehicle without a partner. Before the common corridor state is realized, the operator commits to a public signaling rule; after observing the state, it generates the message according to that rule. Vehicles know this rule, update their beliefs, and choose whether to join. The operator maximizes expected participation subject to Bayesian consistency and individual incentives. This observable formation-stage objective is evaluated alongside fuel savings and monetary utility. Position assignment is fixed, and monetary transfers are absent from the modeled participation stage. Behavioral stability here means that each vehicle's participation strategy is a best response given its information and the strategies of others; subsequent coalition stability and position renegotiation are separate implementation questions.

Transportation research provides substantial evidence that the effects of information depend on travelers' incentives and their equilibrium responses. Table~\ref{tab:transport-information-design} compares representative information-design and incentive-compatible recommendation applications. The table highlights three distinctions. First, information design already addresses many interacting receivers, including finite groups of drivers and infrastructure operators; extending persuasion to multiple agents is therefore not sufficient as a novelty claim. Second, sender objectives range from congestion reduction to safety and infrastructure welfare, and recent charging applications already incorporate heterogeneous driver preferences. Third, the information asymmetry must distinguish an informed sender's common-state knowledge, receiver-specific preferences, and the messages that receivers obtain. For example, \citet{wang2026charging} find full public revelation optimal in their stylized one-station, two-state model with linear costs, while a cutoff-based private design can improve system outcomes. Such results motivate examining how the transportation interaction and the source of private heterogeneity change the value of disclosure.

\begin{table}[htbp]
\centering
\renewcommand{\baselinestretch}{1}\footnotesize\selectfont
\renewcommand{\arraystretch}{1.12}
\setlength{\tabcolsep}{3pt}
\caption{Representative applications of information design and incentive-compatible recommendations in transportation, and the positioning of this paper.}\label{tab:transport-information-design}
\begin{tabular}{@{}>{\raggedright\arraybackslash}p{0.18\linewidth}>{\raggedright\arraybackslash}p{0.185\linewidth}>{\raggedright\arraybackslash}p{0.17\linewidth}>{\raggedright\arraybackslash}p{0.14\linewidth}>{\raggedright\arraybackslash}p{0.26\linewidth}@{}}
\hline
\textbf{Study} & \textbf{Problem scenario} & \textbf{Sender objective} & \textbf{Number of receivers (agents)} & \textbf{Information asymmetry}\\\hline
\citet{wu2019information}\newline (Allerton) & Route choice under uncertain network conditions & Minimize expected spillover onto a protected route & Continuum of travelers & Authority observes the network state; a fixed fraction receives a common noisy signal; others retain the prior.\\[4pt]
\citet{zhu2022information}\newline (TCNS) & Routing with partial information-service participation & Minimize expected total travel latency & Continuum of travelers & Designer conditions messages on the network state; subscribers receive public messages or private recommendations; others use the prior.\\[4pt]
\citet{ning2023robust}\newline (TR-B) & Real-time routing through correlated recommendations & Reduce system travel cost subject to individual incentives & Finite group of individual drivers & Central planner has broader traffic information; drivers receive correlated route guidance rather than choosing independently from local information.\\[4pt]
\citet{gould2023information}\newline (TR-C) & Hazard warnings with partial V2V adoption & Minimize equilibrium accident probability & Continuum of drivers & Noisy hazard detection generates warnings for V2V-equipped vehicles; other drivers receive none; hazard risk depends on behavior.\\[4pt]
\citet{zheng2025government}\newline (TR-B) & Infrastructure adaptation with investment spillovers & Maximize social welfare from adaptation & Finite $N$ infrastructure operators & Government observes the disaster state; operators act on recommendations and a common prior; adaptation costs differ across operators.\\[4pt]
\citet{wang2026charging}\newline (TS) & EV charging under uncertain conditions and congestion & Improve system charging efficiency & Multiple EV drivers with heterogeneous remaining range & Designer knows the charging condition; drivers have range-dependent payoffs; public design is common and private design depends on type.\\[4pt]
\textbf{This paper} & Voluntary platoon formation with partner-dependent fuel savings & Maximize expected platooning participation & Finite $N$ eligible vehicles & Operator observes the common corridor state and broadcasts one public message; vehicles additionally know either private joining costs or noisy private state observations.\\\hline
\end{tabular}
\par\smallskip
% \begin{minipage}{\linewidth}
% \footnotesize\textit{Notes:} The comparison is not an exhaustive review. Sender-generated private recommendations are distinct from receivers' independently acquired private observations. The adaptation study includes heterogeneous costs; heterogeneity alone does not imply that those costs are privately known.
% \end{minipage}
\end{table}

The novelty is to treat public disclosure about a common corridor opportunity as a design instrument for voluntary platoon formation. The operator broadcasts one message, which each self-interested vehicle interprets together with its private information. Joining decisions are interdependent because a vehicle receives no platooning benefit without a partner. Within this common transportation setting, we compare private operating costs and private assessments of downstream conditions. This is distinct from redistributing realized benefits or assigning leaders and partners, and complements those approaches.

The paper makes three contributions:
\begin{enumerate}
\item \emph{A transportation-based information-design framework.} A finite-vehicle coordination game links public messages, private information, and partner-dependent fuel savings. The operator maximizes voluntary participation subject to Bayesian consistency and individual incentives.
\item \emph{A comparison of information environments.} Independent private costs reduce equilibrium responses to each message's posterior mean. Private observations depend on the full posterior and generate a mixture of participant-count distributions. These differences require distinct disclosure formulations and computational methods.
\item \emph{Quantified disclosure gains and policy implications.} Controlled experiments show when optimized disclosure improves participation, why full revelation need not be best, and how costs, observation accuracy, and physical calibration matter. Formation, fuel savings, and utility reveal the limits of participation as a policy objective.
\end{enumerate}
At the reference point, feasible two-message policies support 8.74\% participation with private costs and 51.83\% with private observations, compared with zero without disclosure. Full disclosure supports only 5.15\% and 39.27\%. Gains are concentrated in intermediate opportunities: unfavorable conditions cannot sustain economically attractive formation, while favorable conditions may already support near-full participation. In the reference observation case, the participation-maximizing policy increases gross fuel savings but lowers expected utility from \$0.087 to \$0.076 per eligible vehicle relative to full disclosure.

The framework draws on information design \citep{bergemann2019information,kolotilin2018optimal} and coordination under public and private information \citep{morris2002social,inostroza2025adversarial}. Multiple receivers and threshold strategies are established ideas; the contribution here is their application to partner-dependent platoon formation and the comparison of its two private-information regimes. We focus on modeling, computation, and policy interpretation. Detailed existence proofs and numerical-certification procedures are omitted; participation gains refer to supported symmetric equilibria, not guaranteed adoption.

Section~\ref{sec:model} introduces the model, Section~\ref{sec:equilibrium} explains equilibrium behavior, and Section~\ref{sec:design} formulates disclosure design. Sections~\ref{sec:numerical} and~\ref{sec:discussion} present the numerical evidence and policy implications.

\FloatBarrier
\section{Model: Common Environment and Information Regimes}\label{sec:model}

\subsection*{Model overview}
\begin{figure}[tbp]
\centering
\includegraphics[width=\linewidth]{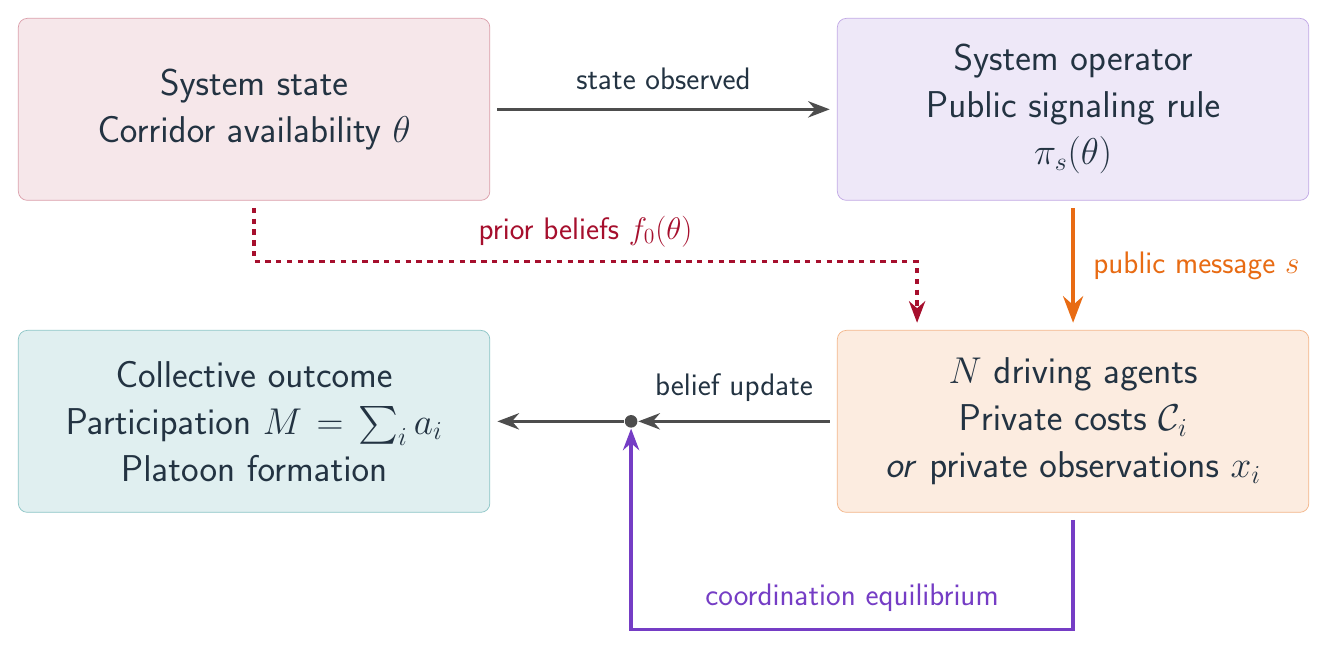}
\caption{Conceptual framework for public information design and voluntary platoon formation. The operator observes the corridor state and broadcasts a public message. Vehicles combine it with prior beliefs and either private costs or private observations, update beliefs, and make interdependent joining decisions. Their equilibrium actions determine participation and formation. The dashed arrow represents prior uncertainty, not direct observation of the realized state.}\label{fig:conceptual-framework}
\end{figure}

Figure~\ref{fig:conceptual-framework} summarizes the agents and information flows in the system. The operator commits to a public signaling rule, observes the common corridor state, and broadcasts a message under that rule. Driving agents initially hold prior beliefs about the state and additionally know either their own joining costs or private observations of corridor conditions. The public message updates their beliefs, including expectations about other vehicles' willingness to join. Each vehicle then chooses whether to participate, and these interdependent decisions determine collective participation and platoon formation. The operator designs disclosure with these behavioral responses in mind. Table~\ref{tab:key-notation} lists the principal notation.
\begin{table}[tbp]\centering\small
\caption{Principal notation. Monetary quantities use calligraphic cost symbols; the additive illustration uses normalized costs.}\label{tab:key-notation}
\begin{tabular}{p{.25\linewidth}p{.67\linewidth}}\hline
Symbol & Meaning\\\hline
$N$, $a_i$, $M$, $K_{-i}$ & Eligible vehicles, joining decision, total joiners, and other joiners.\\
$\theta$, $\overline L$, $L(\theta)$ & Corridor availability, common length, and usable distance $\overline L\theta$.\\
$\mathcal A$, $\overline\delta_n$, $\mathcal R_N$ & Fuel-value scale, average drag reduction, and expected reduction under random participation.\\
$\mathcal C_i$, $\mathcal c$, $\mathcal G$ & Monetary private cost, common cost, and private-cost CDF.\\
$\pi_s$, $P_s$, $f_s$, $m_s$ & Public signaling rule, message mass, posterior density, and posterior mean.\\
$x_i$, $\sigma$, $t_s$, $p_s(\theta;t_s)$ & Private observation, noise scale, observation cutoff, and conditional joining probability.\\
$q_s$, $Q^+(m)$ & Cost-regime participation and greatest supported response at mean $m$.\\
$C_i$, $c$, $b$, $G$ & Normalized costs, complementarity, and cost CDF in the additive illustration.\\
$J$, $w_{ks}$, $\overline S$ & Expected participation fraction, joint cell/message mass, and message budget.\\\hline
\end{tabular}\end{table}

\subsection{Participation stage and transportation payoff}\label{sec:transport-payoff}
Consider $N\geq2$ vehicles eligible to assemble at a specified rendezvous location and departure time, with a common downstream route segment. Vehicle $i$ chooses $a_i\in\{0,1\}$, where $a_i=1$ commits it to the assembly protocol. Let
\begin{equation}
 K_{-i}=\sum_{j\ne i}a_j,\qquad q_{-i}=\frac{K_{-i}}{N-1}.
\end{equation}
Conditional on joining, its prospective platoon size is $n=K_{-i}+1$. Route compatibility, feasible arrival times, and an admissible car-following controller are checked before vehicles enter this candidate set. We specialize the physical interpretation to comparable heavy-duty vehicles, for which experimental fuel-saving evidence is available; other vehicle classes require separate calibration.

Let $\overline L$ denote the known length of the candidate vehicles' common downstream corridor, measured from the rendezvous. Only part of this corridor may support sustained close-following operation: congestion, merging traffic, or restrictions on an eligible lane can require vehicles to increase their spacing or temporarily travel independently. We define $\theta\in[0,1]$ as the fraction of this corridor available for the prescribed platooning operation, and
\begin{equation}\label{eq:effective-distance}
 L(\theta)=\overline L\theta
\end{equation}
as the corresponding effective platooning distance. Thus $\theta$ is a dimensionless measure of the opportunity to platoon, while $L(\theta)$ is a distance over which the fuel-saving mechanism can operate. For example, if a 50-km common corridor contains 30 km suitable for the prescribed operation, then $\theta=0.6$ and $L(\theta)=30$ km. This is a distance fraction, not the probability that other vehicles join.

The operator may aggregate downstream infrastructure measurements and connected-vehicle reports that individual vehicles do not possess before committing to assembly. Vehicles instead share a prior $f_0$ and may have private local assessments. Exact operator observation is an idealization of this information advantage. The state is exogenous to joining decisions and measures available distance, not the probability of obtaining a partner; a favorable corridor alone gives no benefit to a vehicle joining in isolation.

A transportation-based monetary payoff, relative to traveling independently, is
\begin{equation}\label{eq:physical-payoff}
 \widetilde u_i
 =a_i\left[p_f\overline L\theta\,\kappa(v)\overline\delta_{K_{-i}+1}(d)
             -\mathcal C_i\right],
 \qquad
 \mathcal C_i=p_f\Delta F_i^{\mathrm{access}}+w_i\Delta T_i+\chi_i.
\end{equation}
Here $p_f$ is the fuel price, $\kappa(v)$ is the solo aerodynamic fuel use per unit distance, $\overline\delta_n(d)$ is the expected proportional reduction of aerodynamic drag for a participant in an $n$-vehicle platoon, $\Delta F_i^{\mathrm{access}}$ is the incremental fuel needed to reach and assemble at the rendezvous, $w_i$ is the value of time, $\Delta T_i$ is the incremental trip time, and $\chi_i$ is a residual maneuvering or scheduling cost. Fuel and time differences are measured against a solo trip over the same origin--destination movement. They may be signed; nonnegative net joining costs are an additional assumption in the analytical cost examples. Set $\overline\delta_1(d)=0$: a vehicle that commits but finds no partner receives no aerodynamic saving, while any assembly cost it has incurred remains a cost.

\paragraph{Physical interpretation and evidence.}
Two-truck tests report total fuel reductions of 2.7--5.3\% for leaders and 2.8--9.7\% for followers, depending on speed, gap, and mass \citep{lammert2014effect}; four-truck tests also show position dependence \citep{stegner2021experimental}. These motivate bounded, position-dependent savings rather than a universal participation bonus. Cruise fuel benefits are linked to the longitudinal force balance
\begin{equation}\label{eq:traction-model}
\begin{aligned}
 F_i^{\mathrm{tr}}(t)
 &=m_i\dot v_i(t)+m_i g_0\sin\gamma(t)
   +m_i g_0c_r\cos\gamma(t)\\
 &\quad+\tfrac12\rho A_iC_{D,i}
       [1-\delta_{r,n}(d_i(t))]v_i(t)^2,\\
 \dot F_i(t)&\simeq f_{\mathrm{idle},i}
       +\frac{[F_i^{\mathrm{tr}}(t)v_i(t)]_+}{\eta_i H_f}.
\end{aligned}
\end{equation}
where $m_i$ is mass, $g_0$ gravity, $\gamma$ grade, $c_r$ rolling resistance, $\rho$ air density, $A_iC_{D,i}$ drag area, $\delta_{r,n}$ position-dependent drag reduction, $\eta_i$ efficiency, and $H_f$ fuel energy per unit volume. The notation $[z]_+=\max\{z,0\}$ excludes negative traction power from fuel use. At constant speed with positive traction,
\begin{equation}\label{eq:aerodynamic-fuel}
\kappa(v)=\frac{\rho AC_Dv^2}{2\eta H_f}.
\end{equation}
Grade and coordinated speed affect platooning fuel use \citep{turri2017cooperative}, motivating the trajectory checks in Section~\ref{sec:numerical-validation}. Access costs compare catch-up and waiting with solo travel over the same distance, consistent with \citet{liang2013catchup}. Access fuel is deducted separately from downstream savings. Safe spacing and a feasible assembly protocol are prerequisites for the eligible cohort, rather than benefits traded against fuel savings.

\paragraph{Position assignment and diminishing benefits.}
To retain ex ante symmetry, assume that feasible participants are assigned positions by an exchangeable rule after commitment; for example, an assembly facility can use a uniform ordering lottery. Then
\begin{equation}\label{eq:position-average}
 \overline\delta_n(d)=\frac1n\sum_{r=1}^n\delta_{r,n}(d).
\end{equation}
This is an expectation over physical positions, not a monetary transfer. It requires a feasible ordering protocol; fixed road order would instead require vehicle-specific positional payoffs. A parsimonious calibration family is
\begin{equation}\label{eq:saturating-drag}
 \overline\delta_n(d)=
 \begin{cases}
 0,&n=1,\\
 \dfrac{\delta_L(d)+(n-1)\delta_F(d)}{n},&n\geq2,
 \end{cases}
 \qquad 0\leq\delta_L(d)<\delta_F(d)<1.
\end{equation}
This family averages one leader and $n-1$ followers and approaches a bounded follower benefit as size increases. Wind-tunnel evidence documents position effects and drag approaching a plateau in long platoons \citep{robertson2019experimental}. The two-coefficient formula is our simplifying specification, not an empirical law; measured position-specific coefficients can replace it. Ex ante willingness to accept the assignment does not establish willingness to remain in every realized position after commitment.

\subsection{Relation to the tractable additive payoff}\label{sec:payoff-approximation}
Define $\mathcal A=p_f\overline L\kappa(v)$ and $R_N(q)=\overline\delta_{1+(N-1)q}(d)$, using a smooth local interpolation between feasible integer platoon sizes. Equation~\eqref{eq:physical-payoff} has benefit $\mathcal A\theta R_N(q_{-i})$: state quality and participation interact multiplicatively. To connect this physical model to the analytical benchmark below, expand around an interior operating point $(\theta_*,q_*)$ with $R_N(q_*)>0$:
\begin{equation}\label{eq:payoff-linearization}
 \theta R_N(q)\simeq R_N(q_*)\theta
       +\theta_*R_N'(q_*)(q-q_*).
\end{equation}
Dividing monetary utility by $\mathcal A R_N(q_*)$ yields the additive approximation
\begin{equation}\label{eq:general-payoff}
 u_i=a_i[\theta-C_i+B_N(K_{-i})],
\end{equation}
where, for the linear baseline,
\begin{equation}\label{eq:linear-payoff}
 B_N(K_{-i})=bq_{-i},\qquad
 b=\frac{\theta_*R_N'(q_*)}{R_N(q_*)},\qquad
 C_i=\frac{\mathcal C_i}{\mathcal A R_N(q_*)}+bq_*.
\end{equation}
For a common monetary cost $\mathcal c$, the corresponding normalized cost is $c=\mathcal c/[\mathcal A R_N(q_*)]+bq_*$. An operating region with $R_N'(q_*)>0$ gives strategic complementarities. Thus $b$ is tied to the marginal expected drag benefit of additional participation, and $C_i$ is a normalized effective cost, including the expansion's intercept adjustment. It is not the monetary cost $\mathcal C_i$ itself.

The uniform-cost illustration in Section~\ref{sec:design} uses this additive approximation. The main equilibrium and disclosure formulations instead use the physical payoff. The approximation need not preserve zero fuel-saving benefit for an isolated participant, so its predictions are checked against physical incentives in Section~\ref{sec:numerical-validation}.

\subsection{Operator, public information, and timing}
The operator observes $\theta$ and commits to a public signaling rule with a fixed finite message set $S$:
\begin{equation}\label{eq:signal-feasibility}
 \pi_s(\theta)=\Prb(s\mid\theta),\qquad
 \pi_s(\theta)\geq0,\qquad \sum_{s\in S}\pi_s(\theta)=1.
\end{equation}
The timing is:
\begin{enumerate}
 \item The operator commits to a publicly known signaling rule.
 \item The state and private vehicle types or observations are realized.
 \item The operator observes the state and generates a common message.
 \item Vehicles observe the message and their private information, then choose simultaneously.
 \item Participation and payoffs are realized.
\end{enumerate}
The operator's randomization is independent of private types or observation noises conditional on the state. For a message of positive probability, define
\begin{equation}\label{eq:public-posterior}
 P_s=\int_0^1f_0(\theta)\pi_s(\theta)d\theta,\quad
 f_s(\theta)=\frac{f_0(\theta)\pi_s(\theta)}{P_s},\quad
 m_s=\int_0^1\theta f_s(\theta)d\theta.
\end{equation}
Bayes plausibility is the distributional identity
\begin{equation}\label{eq:bayes-plausibility}
 \sum_{s\in S}P_sf_s(\theta)=f_0(\theta).
\end{equation}
In particular, $\sum_sP_sm_s=\E[\theta]$, although this mean identity alone does not characterize all implementable posterior distributions.

\subsection{Two private-information regimes}
\paragraph{Private joining costs.}
Vehicles may face different waiting times, catch-up fuel use, and delivery constraints. These burdens are known to the vehicle or carrier but need not be observed by the operator; preference-based platooning research likewise incorporates fuel and schedule tradeoffs \citep{barua2023participation}. Monetary costs $\mathcal C_i$ have continuous CDF $\mathcal G$, independently across vehicles and independently of $\theta$. Each vehicle knows its own cost, while all share the public posterior. The operator knows the distribution, not individual realizations.

\paragraph{Private observations.}
Vehicles may instead differ in their assessments of the same corridor because their local observations cover different locations or have different accuracy. Holding monetary cost fixed at $\mathcal c$, each vehicle observes
\begin{equation}\label{eq:private-observation}
 x_i=\theta+\varepsilon_i,\qquad \varepsilon_i\overset{\mathrm{iid}}{\sim}\mathcal N(0,\sigma^2),
\end{equation}
with errors independent of the state. The operator does not observe $x_i$. This Bayesian coordination game has the private-observation structure of a global game: a vehicle uses its evidence to infer both the opportunity and others' likely decisions. Conditional independence of errors does not imply unconditional independence of actions. Gaussian observations are not clipped to $[0,1]$.

\subsection{Operator objective and incentive requirement}
The operator maximizes the expected participation fraction,
\begin{equation}\label{eq:objective-common}
 J=\E\left[\frac{1}{N}\sum_{i=1}^N a_i\right].
\end{equation}
The platform can record accepted commitments directly, whereas fuel and welfare objectives require additional physical or private monetary valuations. Participation maximization has a precedent in preference-based platoon matching \citep{barua2023participation}. Here a commitment may still fail to produce a partner, so formation probability, vehicle-distance, fuel savings, and utility are evaluated alongside $J$. The objective concerns uptake among a fixed eligible cohort, not comprehensive social welfare.
\begin{definition}[Persuasive participation strategy]
A signaling scheme and induced strategy profile are persuasive if, conditional on each vehicle's private information and the public message, its prescribed action maximizes its expected utility given the other vehicles' strategies.
\end{definition}
This is an interim no-deviation requirement. It does not require every realized payoff to be nonnegative, equalize benefits, or establish post-formation position stability.

\FloatBarrier
\section{Equilibrium Participation and Formation}\label{sec:equilibrium}
\subsection{Private costs and private observations}\label{sec:physical-equilibrium}
For independently participating opponents, the expected drag reduction is
\begin{equation}\label{eq:expected-drag}
\mathcal R_N(p)=\sum_{k=0}^{N-1}\overline\delta_{k+1}(d)\binom{N-1}{k}p^k(1-p)^{N-1-k}.
\end{equation}
This finite-population expression retains the chance of finding no partner. Let $\Phi$ denote the standard normal CDF and $\varphi_\sigma$ the Gaussian density with standard deviation $\sigma$. Assume the drag coefficients are nonnegative and nondecreasing in platoon size, with $\overline\delta_1=0$.

\begin{proposition}[Participation under the physical payoff]\label{prop:physical-equilibrium}
After a positive-probability message $s$:
\begin{enumerate}
\item With independent private costs, symmetric participation is characterized by
\begin{equation}\label{eq:physical-cost-equilibrium}
q_s=\mathcal G(\mathcal A m_s\mathcal R_N(q_s)),\qquad
 a_i=\ind\{\mathcal C_i\leq\mathcal A m_s\mathcal R_N(q_s)\}.
\end{equation}
Conditional on $s$, $M\sim\operatorname{Binomial}(N,q_s)$.
\item With private observations and common cost $\mathcal c$, a finite symmetric threshold $a_i=\ind\{x_i\geq t_s\}$ is an equilibrium exactly when
\begin{equation}\label{eq:physical-observation-equilibrium}
\int_0^1[\mathcal A\theta\mathcal R_N(p_s(\theta;t_s))-\mathcal c]f(\theta\mid t_s,s)d\theta=0,
\end{equation}
where
\begin{equation}\label{eq:observation-participation}
p_s(\theta;t)=\Phi((\theta-t)/\sigma),\qquad
f(\theta\mid x,s)=\frac{f_s(\theta)\varphi_\sigma(x-\theta)}{\int_0^1f_s(z)\varphi_\sigma(x-z)dz}.
\end{equation}
Conditional on $(\theta,s)$, $M\sim\operatorname{Binomial}(N,p_s(\theta;t_s))$.
\end{enumerate}
\end{proposition}
\paragraph{Economic explanation.}
With private costs, the public message changes the common expected distance. Vehicles then sort by their own joining burdens, and the fixed point ensures their participation expectations are mutually consistent. With private observations, a favorable $x_i$ raises expected savings and the expected number of willing partners. For any fixed opponents' threshold, the joining gain is nondecreasing in $x_i$, so indifference at $t_s$ supports joining above it and staying out below it. These best-response arguments explain the participation conditions in Proposition~\ref{prop:physical-equilibrium}.

\subsection{Why the information environment matters}
\begin{proposition}[Information structure and participant-count risk]\label{prop:information-structure}
Consider either the additive payoff or the separable physical payoff with independent private costs or conditionally independent Gaussian observation errors.
\begin{enumerate}
 \item With independent private costs, two public posteriors with the same mean have the same symmetric cost-cutoff equilibrium correspondence. The induced count law is binomial for any chosen fixed point.
 \item With private Gaussian observations, equal public posterior means need not produce the same symmetric equilibrium correspondence. This failure occurs even for two messages of one implementable public signaling scheme.
 \item For a finite observation threshold, set $\bar q_s=\E[p_s(\theta;t_s)\mid s]$ and $v_s=\operatorname{Var}(p_s(\theta;t_s)\mid s)$. Then, for distinct vehicles $i,j$,
 \begin{equation}\label{eq:count-risk}
 \operatorname{Cov}(a_i,a_j\mid s)=v_s,\qquad
 \operatorname{Var}(M\mid s)=N\bar q_s(1-\bar q_s)+N(N-1)v_s.
 \end{equation}
 For a nondegenerate public posterior and finite threshold, $v_s>0$. Independent private costs with participation $\bar q_s$ instead give variance $N\bar q_s(1-\bar q_s)$.
\end{enumerate}
\end{proposition}
\paragraph{Explanation and a same-mean example.}
The cost fixed point depends on $f_s$ only through $m_s$, proving posterior-mean sufficiency for that equilibrium correspondence. This reduction fails for observations because private evidence is combined with the entire public posterior. For a concrete physical example, take $f_A=U[0.5,0.7]$ and $f_B=U[0.2,1]$, both with mean $0.6$, and set $\mathcal c=0.4\mathcal A\overline\delta_N$. Always joining is an equilibrium under $A$, but sufficiently unfavorable observations make it unprofitable under $B$. These posteriors are implementable: choose $f_0=(f_A+f_B)/2$ and $\pi_s=f_s/(f_A+f_B)$, yielding equal message masses. Thus equal public means can support different participation behavior.

The count-risk identities follow by conditioning on the common state: $\E[a_ia_j\mid s]=\E[p_s^2\mid s]$. They quantify a formation consequence rather than a universal welfare ranking. With two vehicles, formation probability is $\bar q_s^2+v_s$ under private observations, compared with $\bar q_s^2$ under independent costs at the same mean participation. For larger cohorts, the full binomial mixture determines formation probability. Private observations can concentrate commitments in favorable states; average uptake alone misses this dependence.

\subsection{Coordination and the interpretation of disclosure gains}
\begin{corollary}[A formation-specific implementation limit]\label{cor:physical-implementation}
With strictly positive joining costs and no guaranteed partner, everyone staying out is an equilibrium after every public message. Consequently, worst-equilibrium participation is zero under every signaling policy. If every joining cost exceeds $\mathcal A\overline\delta_N$, staying out is strictly dominant.
\end{corollary}
The argument is immediate: a lone participant earns no aerodynamic saving but incurs its joining cost. Favorable information therefore cannot by itself eliminate universal refusal. We optimize supported participation within symmetric cost-cutoff and observation-threshold equilibria, including constant actions. The same favorable-selection convention is applied to every policy comparison. Threshold strategies are best responses, not commands imposed by the operator; no uniqueness or convergence of equilibrium selection is assumed.

\FloatBarrier
\section{Public Information Design}\label{sec:design}
The operator designs what to disclose, anticipating the participation responses above. A message must be generated by the committed rule and every vehicle's action must be optimal given its public and private information. We jointly optimize disclosure and a supported equilibrium; this differs from guaranteeing the same outcome in every equilibrium.

\subsection{Private costs: allocate posterior means}
The physical-payoff design problem is
\begin{equation}\label{eq:cost-design}
\begin{aligned}
\sup_{\pi,P,m,q}\quad &\sum_{s\in S}P_sq_s\\
\text{subject to}\quad &\pi_s(\theta)\geq0,\quad\sum_s\pi_s(\theta)=1,\\
&P_s=\int f_0(\theta)\pi_s(\theta)d\theta,\quad
 P_sm_s=\int\theta f_0(\theta)\pi_s(\theta)d\theta,\\
&q_s=\mathcal G(\mathcal A m_s\mathcal R_N(q_s)),\quad q_s\in[0,1]\quad(P_s>0).
\end{aligned}
\end{equation}
For supported-participation maximization, retain the greatest fixed point $Q^+(m_s)$. This response is nondecreasing in the posterior mean. The operator therefore allocates state probability across messages to create attractive posterior opportunities while preserving Bayesian consistency.

\subsection{Private observations: account for posterior-dependent incentives}
Expected participation is
\begin{equation}\label{eq:observation-objective}
 J(\pi,t)=\sum_s\int_0^1f_0(\theta)\pi_s(\theta)p_s(\theta;t_s)d\theta.
\end{equation}
For a finite threshold, define the unnormalized conditional joining gain
\begin{equation}\label{eq:unnormalized-gain}
D_s(x;\pi,t_s)=\int_0^1f_0(\theta)\pi_s(\theta)\varphi_\sigma(x-\theta)
[\mathcal A\theta\mathcal R_N(p_s(\theta;t_s))-\mathcal c]d\theta.
\end{equation}
Persuasiveness requires $D_s(x)\geq0$ above $t_s$ and $D_s(x)\leq0$ below it. Monotonicity of the normalized joining gain reduces these conditions to $D_s(t_s)=0$. Hence the observation design maximizes~\eqref{eq:observation-objective} subject to signal feasibility and these message-specific indifference constraints. Always-join and always-out messages are checked separately. In contrast with the cost regime, preserving only posterior means would discard information needed for these constraints.

\subsection{Computational solution}\label{sec:algorithms}
Partition the state space into cells $I_k$ with prior masses $\mu_k$ and conditional means $\theta_k$. A cellwise-constant signal is represented by
\begin{equation}\label{eq:algorithm-mass}
w_{ks}=\Prb(\theta\in I_k,s),\qquad w_{ks}\geq0,\quad\sum_s w_{ks}=\mu_k,
\qquad \pi_s(\theta)=w_{ks}/\mu_k\quad(\theta\in I_k).
\end{equation}
Message means satisfy $P_sm_s=\sum_k\theta_kw_{ks}$. This representation enforces Bayes plausibility directly, with no loss of the continuous state within a cell during observation-likelihood integration.

\begin{algorithm}[tbp]
\caption{Private-cost disclosure by posterior-mean allocation}\label{alg:private-cost}
\begin{algorithmic}[1]
\Require Prior, physical payoff, cost CDF, state cells, message budget, and search budget.
\State Initialize with feasible no-disclosure and fixed coarse-disclosure policies.
\State Partition the possible message posterior means into intervals.
\For{each interval configuration explored}
\State Solve a linear allocation relaxation using $Q^+$ at upper mean endpoints.
\State Recover actual message means and recompute their greatest cost fixed points.
\State Verify cost cutoffs and retain the best supported policy.
\State Refine promising mean intervals using their relaxation upper bounds.
\EndFor
\State \Return signal probabilities, participation, incentive residuals, and remaining numerical gap.
\end{algorithmic}\end{algorithm}
For two messages, branch and bound searches their posterior means. For a larger alphabet, a mean-interval allocation relaxation gives an upper bound; merging recovered allocations and recomputing actual fixed points supplies feasible policies. The upper relaxation may permit more messages than the recovered policy and is reported accordingly.

\begin{algorithm}[tbp]
\caption{Private-observation disclosure by threshold search}\label{alg:private-observation}
\begin{algorithmic}[1]
\Require Prior, physical payoff, common cost, noise scale, state cells, message and search budgets.
\State Include admissible constant-action messages and initialize feasible baseline policies.
\For{each candidate vector of finite thresholds}
\State Compute cell-integrated participation and indifference coefficients.
\State Solve the linear allocation problem subject to message indifference and state marginals.
\State Check conditional incentive residuals and retain the best supported policy.
\EndFor
\State Use interval-threshold relaxations to bound unexplored threshold regions.
\State Refine thresholds or select a budgeted threshold library for larger alphabets.
\State \Return signal probabilities, thresholds, participation, and remaining numerical gap.
\end{algorithmic}\end{algorithm}
At fixed thresholds, both participation and the unnormalized indifference constraints are linear in $w$. Threshold search is therefore combined with LPs, while four-message candidate selection uses a library MILP and subsequent refinement. Constant-action columns retain their support restrictions. Bounds concern the specified state cells, message budget, and finite-threshold range. Implementation uses floating-point arithmetic and independent incentive and integration checks; it does not provide an interval-arithmetic certificate or establish equilibrium selection. Settings and achieved gaps are reported in Section~\ref{sec:numerical}.

\subsection{An additive illustration of why cost distributions matter}
The normalized approximation~\eqref{eq:general-payoff}--\eqref{eq:linear-payoff} gives $q_s=G(m_s+bq_s)$, where $G$ is the CDF of normalized costs and $g$ its density when it exists. If $b\sup g<1$, the response is unique; denote it by $Q(m_s)$. For $C_i\sim U[\underline c,\overline c]$ and $0<b<\overline c-\underline c$,
\begin{equation}\label{eq:uniform-cost-response}
Q(m)=\begin{cases}0,&m\leq\underline c,\\
(m-\underline c)/(\overline c-\underline c-b),&\underline c<m<\overline c-b,\\
1,&m\geq\overline c-b.
\end{cases}
\end{equation}
\begin{proposition}[An uninformative policy can be optimal]\label{prop:no-disclosure}
For $C_i\sim U[0,\overline c]$ with $0<b<\overline c$, no additional disclosure maximizes participation in the additive model.
\end{proposition}
Here $Q(m)=\min\{m/(\overline c-b),1\}$ is concave. Bayesian consistency and Jensen's inequality give $\sum_sP_sQ(m_s)\leq Q(\E[\theta])$, attained by a constant message.

\begin{example}[A positive minimum cost makes informative disclosure useful]\label{ex:positive-cost}
Let $\theta\sim U[0,1]$, $C_i\sim U[0.6,1.2]$, and $b=0.2$. Without additional information, $m=0.5$ and $Q(m)=0$. Send $H$ when $\theta\geq0.6$ and $L$ otherwise. Then $P_H=0.4$, $m_H=0.8$, $Q(m_H)=0.5$, while $P_L=0.6$, $m_L=0.3$, and $Q(m_L)=0$. Expected participation is $0.2$. The high-message cutoff is $\tau_H=0.9$, and all cost types follow their best responses. This is a feasible improvement with a unique symmetric fixed point, not an assertion that this signal is globally optimal.
\end{example}
These examples establish that the value of disclosure depends on the cost distribution, without implying a general binary-signal optimum or transferring the additive result to physical formation incentives. The experiments compare both payoff specifications and use the physical payoff for transportation conclusions.

\FloatBarrier
\section{Numerical Analysis: Disclosure, Heterogeneity, and Platoon Formation}\label{sec:numerical}
The numerical study addresses three questions: how signaling changes voluntary platoon formation; how private joining costs and private observations alter that response; and how much optimized disclosure improves participation relative to a platform providing no additional public information. These experiments supply the evidence for the disclosure-targeting and objective-tradeoff implications stated in the introduction; the equilibrium structure and implementation limit are explained in Section~\ref{sec:equilibrium}. We use a literature-anchored highway truck scenario and controlled parameter changes, followed by physical-payoff and equilibrium-selection checks. Sections~\ref{sec:numerical-baseline-results} and~\ref{sec:numerical-design-results} report the baseline and 39 implemented physical-payoff configurations, each comparing four policies. Section~\ref{sec:numerical-validation} completes the analytical checks, message-budget and state-grid refinements, payoff approximation comparison, parameter sensitivities, and constructed-trajectory validation. These additions are archived separately from the original 39 configurations; no field validation is claimed.

\subsection{Literature-calibrated operating environment}\label{sec:numerical-calibration}
Each decision round contains a fixed cohort of route-compatible heavy-duty vehicles approaching the same rendezvous. The cohort, departure protocol, feasible operating speed, and following gap are identical across the signaling policies compared in that round. The physical payoff~\eqref{eq:physical-payoff} provides the operational interpretation and monetary evaluation. The additive payoff supplies analytical examples. A separate numerical comparison fixes its expansion point and tests whether its designed signals retain their predicted performance when vehicles instead respond to the physical payoff.

The operating envelope is anchored to the two-truck tests of \citet{lammert2014effect}, which covered speeds of 55--70 mph, gaps of 20--75 ft, and gross weights of 65,000 and 80,000 lb. We use the 55-mph, 30-ft, 65,000-lb condition as the reference operating point. Evidence for position-dependent savings in four-truck platoons comes from \citet{stegner2021experimental}; those tests used a different fleet and speed, so they provide an additional plausibility check rather than direct validation of the reference configuration. Catch-up costs follow the trajectory accounting motivated by \citet{liang2013catchup}. Table~\ref{tab:numerical-inputs} distinguishes measured operating conditions, externally grounded valuations, and scenario assumptions.

\begin{table}[tbp]
\centering
\renewcommand{\baselinestretch}{1.05}\small\selectfont
\renewcommand{\arraystretch}{1.15}
\begin{tabular}{p{0.23\linewidth}p{0.29\linewidth}p{0.37\linewidth}}
\hline
Input & Reference and variations & Basis and interpretation\\
\hline
Speed, gap, mass & 55 mph, 30 ft, 65,000 lb (about 88.5 km/h, 9.14 m, 29.5 t) & Tested operating point in \citet{lammert2014effect}; held fixed in the main design grid.\\
Candidate vehicles $N$ & 4; compare 2 and 4; 8 as sensitivity only & Small-platoon experiments motivate the primary sizes; eight-vehicle benefits require extrapolation.\\
Common corridor $\overline L$ & 100 km; 50 and 200 km & Constructed short, reference, and long overlap scenarios; not measured route frequencies.\\
Solo fuel use $e_0$ & 0.30 L/km; 0.25--0.40 L/km & Illustrative calibration range, to be checked against the chosen vehicle and drive cycle.\\
Cruise fuel-saving targets & Leader 4\%, follower 8\% at the reference point & Rounded scenario targets within reported experimental ranges; not universal coefficients.\\
Aerodynamic share $\xi$ & 0.50 & Assumed conversion from total-fuel savings to drag reductions; varying it at fixed fuel-saving targets leaves the constant-speed benefit unchanged.\\
Fuel price $p_f$ & \$1.10/L; \$0.80--1.40/L & Illustrative constant-2023-dollar scenarios, not a claim about current prices.\\
Driver time value $w$ & \$35.70/hour & USDOT Table A-2, 2023 dollars; excludes freight time value \citep{usdot2024bca}.\\
Mean joining cost $\overline C$ & \$1.50; \$0.75 and \$3.00 & Constructed levels; a separate access-trajectory experiment tests their operational interpretation.\\
State prior & $\mathrm{Beta}(6,4)$; mean $\mu_\theta=0.3,0.6,0.9$ & Constructed corridor-availability scenarios, with concentration initially fixed at 10.\\
\hline
\end{tabular}
\caption{Physical calibration and controlled scenario inputs. Monetary assumptions are expressed in constant 2023 dollars. Values marked as scenario assumptions are not empirical estimates.}
\label{tab:numerical-inputs}
\end{table}

Calibration distinguishes total fuel savings from aerodynamic drag reduction. At the reference speed, we set $\kappa(v)=\xi e_0$ and match the illustrative leader and follower fuel-saving fractions $r_L=0.04$ and $r_F=0.08$ through
\begin{equation}\label{eq:numerical-drag-calibration}
 \delta_L=\frac{r_Le_0}{\kappa(v)}=\frac{r_L}{\xi},\qquad
 \delta_F=\frac{r_Fe_0}{\kappa(v)}=\frac{r_F}{\xi}.
\end{equation}
The reference assumptions yield $\delta_L=0.08$ and $\delta_F=0.16$; size-dependent drag benefits follow~\eqref{eq:saturating-drag}. Speed-varying validation profiles retain these drag coefficients and compute fuel consumption from the trajectory model. Separate spacing and weight effects are not estimated. The two-coefficient size formula remains an approximation, especially for the eight-vehicle sensitivity case.

The core cost distribution aggregates excess fuel, delay, and scheduling burden. At the reference time value, two minutes cost about \$1.19 before fuel or scheduling costs, giving the \$1.50 mean an interpretable scale. The separate validation in Section~\ref{sec:numerical-validation} constructs 501 acceleration- and speed-constrained access paths and charges excess fuel relative to the same origin--destination solo trip, plus net additional time. Its costs are generated independently of participation outcomes and are not fitted to obtain a desired disclosure gain. Fuel and time components are retained separately for net-fuel accounting; no generic operating-cost rate is added on top of them. The trajectories are constructed engineering scenarios, rather than proprietary fleet records or new field measurements.

\subsection{Controlled construction of the two information regimes}\label{sec:numerical-types}
For private costs, we use a mean-preserving uniform family for monetary costs,
\begin{equation}\label{eq:numerical-cost-family}
 \mathcal C_i\sim U\left[\overline C(1-\sqrt{3}\nu_C),
                         \overline C(1+\sqrt{3}\nu_C)\right],
 \qquad \nu_C\in\{0.15,0.35,0.55\}.
\end{equation}
Here $\nu_C$ is the coefficient of variation; all three settings have positive lower support and the same mean. The reference is $\nu_C=0.35$. This family isolates dispersion and supports the polynomial root procedure for the physical payoff. A continuous piecewise-linear CDF fitted to feasible access costs supplies a separate shape check against a uniform distribution with matched mean and variance. Homogeneous costs are treated as a benchmark rather than through a continuous cost density.

For private observations, we set the monetary joining cost to the same reference mean, $\mathcal c=\overline C$, and vary
\begin{equation}\label{eq:numerical-noise-family}
 x_i=\theta+\varepsilon_i,\qquad
 \varepsilon_i\sim\mathcal N(0,\sigma^2),\qquad
 \sigma\in\{0.05,0.15,0.30\}.
\end{equation}
The reference is $\sigma=0.15$. For a 100-km corridor these standard deviations correspond to assessments of effective distance with errors of 5, 15, and 30 km. They are controlled accuracy scenarios, not estimated sensor-error parameters. Observations are not clipped to $[0,1]$, preserving the Gaussian likelihood.

We use $\theta\sim\mathrm{Beta}(\mu_\theta\kappa_\theta,(1-\mu_\theta)\kappa_\theta)$, with reference values $\mu_\theta=0.6$ and $\kappa_\theta=10$. Changes in the mean represent low, intermediate, and high availability; changes in $\kappa_\theta\in\{4,10,30\}$ at fixed mean isolate prior uncertainty. Costs are independent of the state, as required by the theory. The two regimes share physical calibration, prior, and mean cost, but their dispersion parameters are not interchangeable. Within-regime policy gains are the primary comparisons; cross-regime levels do not give a welfare ranking at matched informational content.

For additive-model experiments, we map monetary quantities to $(C_i,c,b)$ using~\eqref{eq:linear-payoff}, fixing $(\theta_*,q_*)=(0.6,0.5)$ across policies and information treatments. The interpolation $R_N$ in~\eqref{eq:payoff-linearization} is differentiated before taking the expectation over the number of other joiners; it is distinct from the exact binomial polynomial $\mathcal R_N$. At $N=4$, the chosen interpolation gives $R_4(0.5)=0.128$, $R_4'(0.5)=0.0384$, and $b=0.18$. Keeping these coefficients fixed prevents a policy-dependent change of payoff. The largest monetary cost-response slope in these six additive scenarios is below $0.42$, ensuring a unique additive cost response.

\subsection{Signaling policies and fair baselines}\label{sec:numerical-policies}
Each parameter configuration is evaluated under the following policies, keeping the eligible cohort and all payoff and information primitives fixed:
\begin{enumerate}
 \item \emph{No additional public disclosure (primary baseline).} The operator sends a constant message. Private costs or private observations remain available to vehicles; the baseline does not remove their own information.
 \item \emph{Fixed coarse disclosure.} The operator sends $H$ if $\theta\geq0.5$ and $L$ otherwise, with the cutoff specified before optimization. This tests whether benefits arise from providing some public information rather than from optimizing its content.
 \item \emph{Optimized public disclosure.} The appropriate algorithm uses at most two messages, including the option to leave a message unused. Selected four-message cases assess the value of a richer signal alphabet.
 \item \emph{Full disclosure.} The state $\theta$ is revealed exactly and the equilibrium is recomputed. This is an informational benchmark, not a participation upper bound, and it lies outside the fixed two-message class.
\end{enumerate}
With consistent equilibrium selection and a feasible baseline incumbent, the optimum over at most two messages cannot be worse than no disclosure; a reported numerical decrease indicates unresolved solver error or a changed comparison. Full disclosure can nonetheless yield more or less participation than a restricted optimized policy. The fixed coarse signal should use the same number of available messages as the primary optimized design.

The same selection convention is used across policies: the unique response in contracting additive cases, and greatest participation among verified symmetric equilibria in the physical comparisons. Root-search qualifications are reported separately from incentive residuals. The adverse-selection outcome is zero for every physical policy with strictly positive costs and no guaranteed partner, by Proposition~\ref{prop:physical-equilibrium}. Guaranteed-partner protocols are outside the experiment and would require different payoff and formation accounting.

The original 39 distinct configurations cover the reference point, a cost-level/dispersion grid, an observation-noise/prior-concentration grid, corridor-length/prior-mean combinations, and cohort-size sensitivities. Supplementary runs test numerical resolution, message budgets, physical calibration, additive approximation, access costs, trajectories, and off-design incentives. All predefined cases, including zero-gain cases, are retained.

\subsection{Outcome measures and improvement relative to baseline}\label{sec:numerical-metrics}
Let $J_0$ be expected participation without additional public disclosure and $J_\pi$ the value under policy $\pi$. The primary effect sizes are
\begin{equation}\label{eq:numerical-gains}
 \Delta_{\mathrm{pp}}=100(J_\pi-J_0),\qquad
 \Delta_N=N(J_\pi-J_0),\qquad
 \Delta_{\mathrm{rel}}=100\frac{J_\pi-J_0}{J_0}\quad(J_0>0).
\end{equation}
Absolute percentage-point changes are reported throughout. Expected-count and relative changes follow~\eqref{eq:numerical-gains}; relative gains are undefined when $J_0=0$. The supplementary summary archives these measures wherever a within-scenario no-disclosure comparison is available.

Additional measures connect the participation objective to transportation outcomes:
\begin{enumerate}
 \item The distribution of $M$, the probability $\Prb(M\geq2)$ of having enough participants for a platoon, and the realized-platooning fraction $\E[M\ind\{M\geq2\}/N]$. These distinguish commitments from viable formation.
 \item Expected platooning vehicle-distance $\E[M L(\theta)\ind\{M\geq2\}]$ and mean platoon size conditional on $M\geq2$. Counting vehicle-distance avoids equating a brief formation with a long shared trip.
 \item Net fuel saved relative to the same cohort traveling independently, subtracting access fuel; total monetary utility; and incurred delay. These use the physical payoff and identified access components, rather than participation alone.
 \item Participation by cost quantile or private-observation bin, conditional expected utility of joiners, and the fraction of joiners experiencing a negative realized payoff. Interim obedience does not eliminate realized losses.
\end{enumerate}
In the private-cost regime, $M\mid s$ is binomial with parameter $q_s$. In the observation regime, $M\mid\theta,s$ is binomial with parameter $p_s(\theta;t_s)$; unconditional metrics integrate over the state, preserving the dependence induced by the common opportunity. The fuel-saving measure includes the position-dependent benefit, using the assignment rule already specified in the model.

\subsection{Reproducibility, precision, and presentation}\label{sec:numerical-reproducibility}
Policies are solved on integrated state cells and evaluated with finer within-cell quadrature. The reference state partitions contain 40, 80, and 160 base cells, with additional structural splits. Two-message mean intervals are refined adaptively; the four-message cost upper relaxation uses 3,200 posterior-mean intervals. The observation-threshold search range is $[-2,3]$, with constant-action messages retained explicitly. This range is a computational restriction, not a theoretical bound on all equilibrium thresholds.

The target objective-bound gap is $10^{-3}$ in participation fraction, or 0.1 percentage point; the target conditional incentive residual is $10^{-6}$ in payoff units. Achieved gaps, resource limits, runtimes, roots, cell resolutions, and threshold ranges are archived. A resource-limited result is a feasible numerical incumbent with a bound, rather than a proved unrestricted optimum. Root-search and equilibrium-selection qualifications are separate from sampling confidence intervals.

Table~\ref{tab:computational-settings} consolidates the reference-case settings and archived search times. All finite observation thresholds lie in $[-2,3]$, with constant-action messages handled separately. The main two-message runs allow 6,000 initial and 24,000 refined search nodes; four-message observation library MILPs have a 30-second limit per solve, with additional threshold refinement and upper-bound calculations. These limits are resource budgets, not stopping certificates.

\begin{table}[tbp]
\centering\small\setlength{\tabcolsep}{4pt}
\caption{Computational settings for representative reference-case runs. Cells include structural splits beyond the 80-cell base partition. GL gives Gauss--Legendre nodes per cell for coefficient or outcome evaluation: solve order; verification orders. Cost posterior means are integrated analytically. Gap is the numerical upper bound minus supported incumbent in percentage points. Times are archived seconds for the reported search or refinement stage, excluding prior incumbent searches, simulations, and independent audits.}\label{tab:computational-settings}
\begin{tabular}{lrrrr}
\hline
Regime and message budget & Cells & GL & Gap (pp) & Time (s)\\\hline
Costs, $|S|\leq 2$ & 82 & 8; 16,32 & 0.10 & 0.21\\
Observations, $|S|\leq 2$ & 82 & 8; 16,32 & 0.18 & 97.73\\
Costs, $|S|\leq 4$ & 82 & 16; 32,64 & 0.04 & 5.77\\
Observations, $|S|\leq 4$ & 82 & 16; 32,64 & 0.89 & 42.94\\
\hline
\end{tabular}
\end{table}

Archived runs use Python 3.12.14, NumPy 2.5.3, and SciPy 1.18.1 with HiGHS. Reported times are stored search-stage durations, not standardized hardware benchmarks. Incentive residuals assess approximate obedience; objective gaps assess remaining improvement in the restricted design class; quadrature, cell refinement, and Monte Carlo checks address integration, resolution, and outcome evaluation. These diagnostics do not establish unrestricted optimality or equilibrium selection.

The initial no-disclosure check uses 20,000 independent decision rounds. The 39 main comparisons use 80,000 rounds and common states, public-message uniforms, and private types across policies; policy differences use paired round-level standard errors. The supplementary optimized policies receive separate 80,000-round checks against integrated participation, realized platooning, and vehicle-distance. Random leader assignment also estimates the frequency of negative realized joining payoffs. The independent sampling unit is a decision round, since vehicles share a state and message. Seeds and parameter files are archived, and numerical optimization gaps, Monte Carlo uncertainty, and calibration sensitivity are reported separately.

Results proceed from no-disclosure equilibria to reference signal rules, controlled information treatments, the operating grid, and the supplementary validation. All reported improvements remain conditional on the specified symmetric supported-equilibrium selection. Constructed trajectories and literature-anchored inputs support a controlled numerical study, not field validation.

\subsection{No-disclosure benchmark}\label{sec:numerical-baseline-results}
We first implement the physical-payoff model with a constant public message. This provides a reproducible baseline for subsequent information design, while preserving vehicles' private costs or observations. The reference case uses $N=4$, $\overline L=100$ km, $\theta\sim\mathrm{Beta}(6,4)$, $p_f=\$1.10$/L, solo consumption $e_0=0.30$ L/km, and aerodynamic share $\xi=0.50$. Leader and follower total-fuel reductions of 4\% and 8\% imply drag reductions $\delta_L=0.08$ and $\delta_F=0.16$. These are the literature-anchored experimental inputs above, not parameters estimated from a new traffic dataset. Costs follow~\eqref{eq:numerical-cost-family} with $\nu_C=0.35$, or observations follow~\eqref{eq:numerical-noise-family} with $\sigma=0.15$. We retain the reference mean cost of \$1.50 and separately report the already specified \$0.75 sensitivity in the cost-level comparison.

For this calibration, the exact finite-cohort drag polynomial is $\mathcal R_4(q)=0.36q-0.32q^2+0.10q^3$. Cost equilibria are obtained from the physical fixed point; observation equilibria from message-conditional threshold indifference. Checks use 256/512-point quadrature, refined observation scans on $[-2,3]$, and 20,000 independent decision rounds with common primitive draws. All checked baseline moments lie within 1.63 simulation standard errors of their integrated values. The calculations identify supported equilibria rather than a behavioral learning process.

\begin{table}[htbp]
\centering
\small
\caption{Computed no-disclosure equilibria under the physical payoff. $J_0=\E[M]/4$ is participation and $P_2=\Prb(M\geq2)$ is formation probability. Fuel is gross liters saved per eligible vehicle; utility is dollars per eligible vehicle after joining costs. Observation results concern the symmetric threshold class searched.}\label{tab:baseline-results}
\begin{tabular}{llrrrr}
\hline
Cost & Regime and equilibrium & $100J_0$ & $100P_2$ & Fuel & Utility\\
\hline
\$1.50 & Private costs: $q=0$ & 0 & 0 & 0 & 0\\
\$1.50 & Observations: always out & 0 & 0 & 0 & 0\\
\hline
\$0.75 & Private costs: $q=0$ & 0 & 0 & 0 & 0\\
\$0.75 & Private costs: $q=0.130848$ & 13.08 & 8.57 & 0.049 & 0.008\\
\$0.75 & Private costs: $q=1$ & 100.00 & 100.00 & 1.260 & 0.636\\
\$0.75 & Observations: always out & 0 & 0 & 0 & 0\\
\$0.75 & Observations: $t=0.874664$ & 9.51 & 9.00 & 0.080 & 0.017\\
\$0.75 & Observations: $t=0.049592$ & 99.52 & 99.98 & 1.256 & 0.636\\
\hline
\end{tabular}
\end{table}

\paragraph{Reference outcome and the role of coordination.}
At the \$1.50 reference cost, the cost equation has only the zero-participation solution. The largest computed observation indifference gain is approximately $-\$0.400$, and the threshold search finds no positive-participation equilibrium. Thus the reference case is retained as a zero-participation baseline rather than changing the calibration to obtain a positive result. At the predefined \$0.75 sensitivity, both regimes support high- and low-participation outcomes in addition to everyone staying out (Table~\ref{tab:baseline-results}). Under favorable selection the baseline participation is already 100\% for private costs and 99.52\% for private observations; subsequent information design has at most 0 and 0.48 percentage points of headroom, respectively, in this low-cost case. Under adverse selection both baselines remain zero.

\begin{figure}[htbp]
\centering
\includegraphics[width=\linewidth]{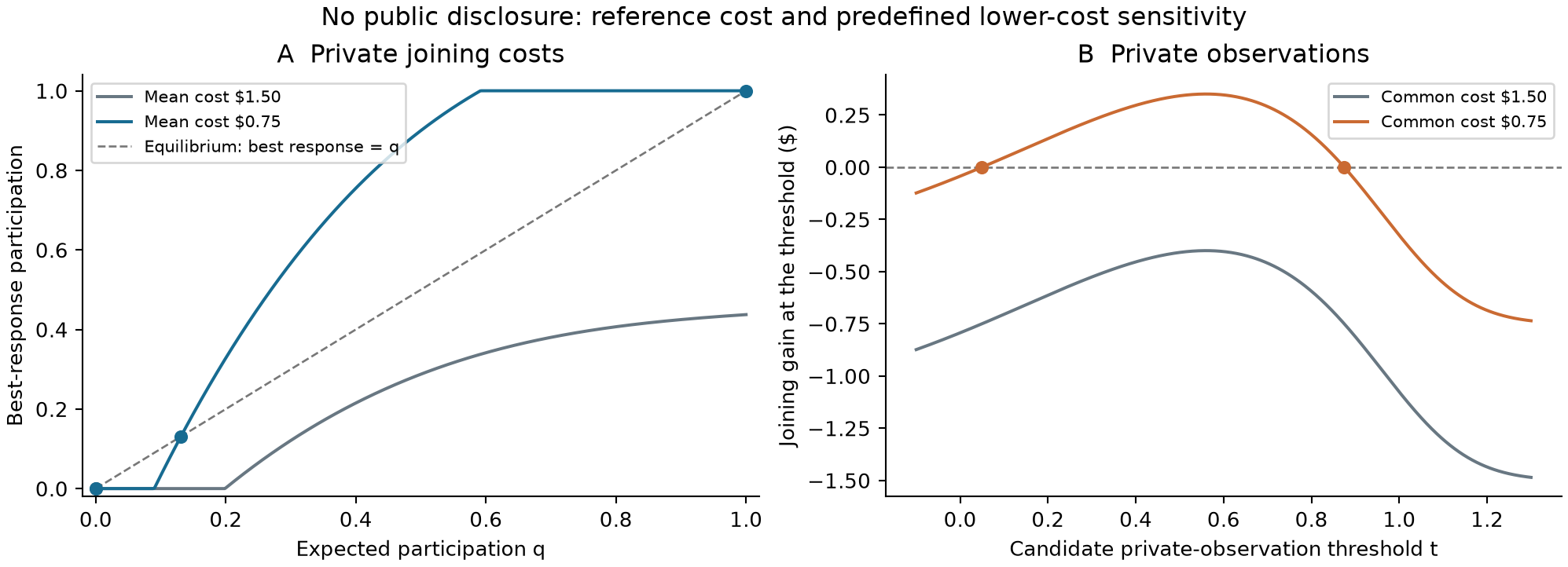}
\caption{No-disclosure equilibrium calculations. Left: intersections of best-response participation with the diagonal identify private-cost equilibria. Right: zeros of the observation indifference gain identify finite thresholds; the always-out equilibrium is additional to the plotted roots. The reference and predefined lower-cost cases are both shown.}\label{fig:baseline-equilibria}
\end{figure}

\begin{figure}[htbp]
\centering
\includegraphics[width=\linewidth]{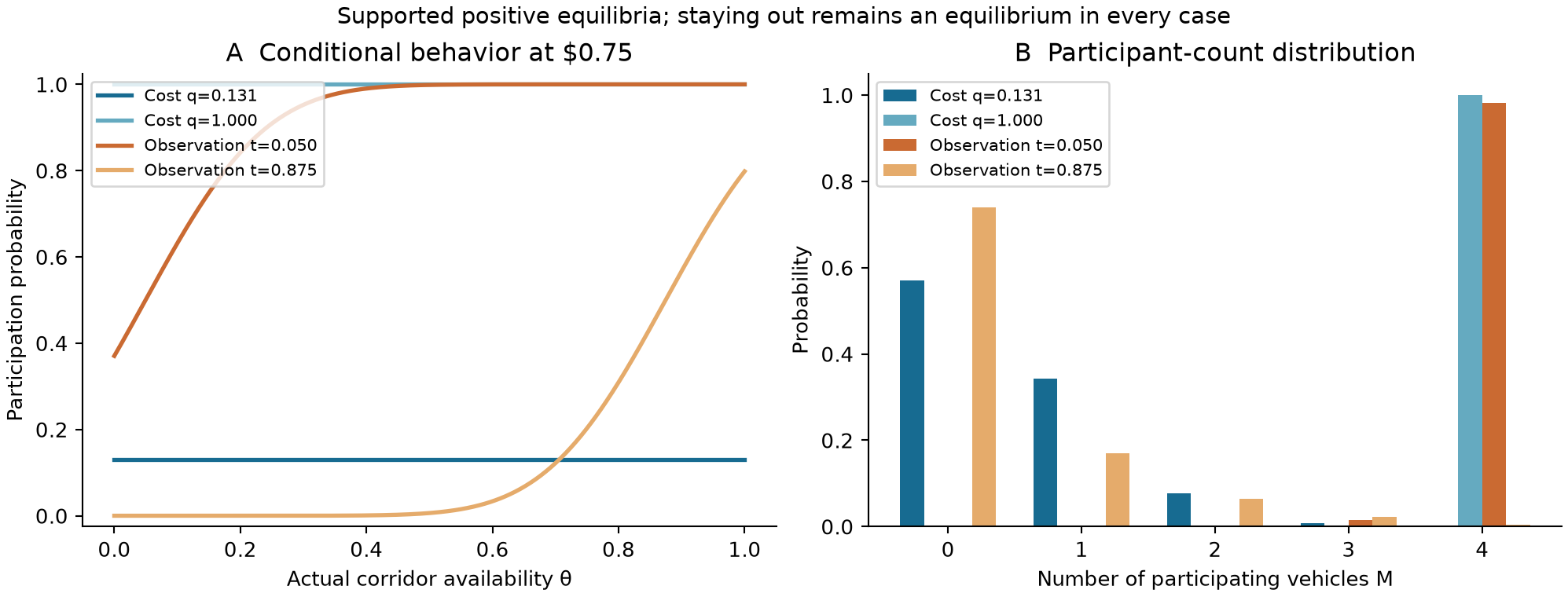}
\caption{Behavior under the positive no-disclosure equilibria at the predefined \$0.75 cost. Left: state-conditional participation. Right: participant-count distributions. Always staying out remains an equilibrium in both regimes and places probability one on $M=0$.}\label{fig:baseline-behavior}
\end{figure}

\paragraph{Different behavior from costs and observations.}
Figure~\ref{fig:baseline-behavior} illustrates the structural mechanism in Proposition~\ref{prop:information-structure}. Independent private costs generate participation that is constant across realized corridor states under no disclosure. Private observations generate participation $p(\theta;t)=\Phi((\theta-t)/\sigma)$ that rises with corridor availability. In the lower-participation equilibria, the observation regime has fewer commitments on average (9.51\% versus 13.08\%), yet a slightly higher formation probability (9.00\% versus 8.57\%) and more gross fuel savings (0.080 versus 0.049 L per eligible vehicle). Joining concentrates in favorable common states, which also correlates vehicles' actions unconditionally. Consequently, its count distribution must integrate state-conditional binomials, rather than use a binomial at average participation. This illustrates a mechanism, not a welfare ranking at matched private information or a prediction that vehicles select the lower-participation branch.

% BEGIN COMPUTED DESIGN RESULTS
\subsection{Disclosure design and policy comparisons}\label{sec:numerical-design-results}
We extend the baseline to 39 distinct controlled parameter configurations, each evaluated under all four disclosure policies. These cover the reference point, the nine cost-level/dispersion cells, the nine observation-noise/prior-concentration cells, the nine corridor-length/prior-mean cells in each regime, the lower-cost observation case, and $N=2$ and $N=8$ sensitivities; shared configurations are run once. We use the physical payoff throughout and at most two public messages. All main comparisons use the highest supported participation found among symmetric cost-cutoff or observation-threshold equilibria, including constant actions, under the same favorable-selection convention for baseline and designed policies. Every policy also admits the all-out equilibrium, giving zero under adverse selection.

\paragraph{Solution quality and scope.}
The stopping target is 0.1 percentage point in participation. Initial searches allow 6,000 nodes; cases retaining larger gaps are rerun with up to 24,000 nodes, and the two reference cases use 80 base state cells in the refined runs. The achieved reference gaps are 0.10 and 0.18 percentage points for private costs and observations, respectively. Across all 39 cases, 31 meet the target and the largest remaining gap is 0.49 percentage points. Reported optimized values are feasible numerical incumbents accompanied by these bounds, not assertions of an exactly attained unrestricted optimum. Bounds concern cellwise-constant signals and, for observations, finite thresholds in the specified interval. Floating-point LP and quadrature calculations are not outward-rounded interval certificates. The reference state-grid comparison and within-cell integration checks assess numerical sensitivity without establishing a bound for arbitrary measurable disclosure.

An independent audit verifies Bayes plausibility, probability bounds, the recovered cost fixed points, and observation incentive signs. Its maximum checked observation incentive violation is below $10^{-6}$ dollars. Reintegrating fixed optimized policies at 32 nodes per cell changes participation by less than $10^{-7}$. We evaluate each policy over 80,000 rounds with common states, public-message uniforms, and private types across policies. Participation differences use paired round-level standard errors; the maximum absolute discrepancy between integrated and simulated participation across the study is 2.36 simulation standard errors. Sampling uncertainty, equilibrium-search qualifications, and optimization gaps remain separate.

\begin{table}[htbp]
\centering\small\setlength{\tabcolsep}{4pt}
\caption{Reference-case policy comparison under favorable symmetric equilibrium selection. $100J$ and $100P_2$ are percentages. Fuel is gross liters saved per eligible vehicle and utility is dollars per eligible vehicle.}\label{tab:design-reference}
\begin{tabular}{llrrrr}
\hline
Regime & Policy & $100J$ & $100P_2$ & Fuel & Utility\\\hline
Private costs & No disclosure & 0.00 & 0.00 & 0.000 & 0.000\\
Private costs & Fixed coarse & 0.00 & 0.00 & 0.000 & 0.000\\
Private costs & Optimized, two messages & 8.74 & 11.99 & 0.118 & 0.039\\
Private costs & Full disclosure & 5.15 & 6.83 & 0.082 & 0.030\\
Private observations & No disclosure & 0.00 & 0.00 & 0.000 & 0.000\\
Private observations & Fixed coarse & 0.00 & 0.00 & 0.000 & 0.000\\
Private observations & Optimized, two messages & 51.83 & 52.86 & 0.776 & 0.076\\
Private observations & Full disclosure & 39.27 & 39.27 & 0.615 & 0.087\\
\hline
\end{tabular}
\end{table}

\paragraph{Research question 1: what does signaling change?}
At the reference point, optimized disclosure raises participation from zero to 8.74\% with private costs and 51.83\% with private observations (Table~\ref{tab:design-reference}). The fixed cutoff at $\theta=0.5$ generates no participation in either regime. Full disclosure produces 5.15\% and 39.27\%, respectively, so merely revealing more information does not maximize participation. The absolute improvements over no disclosure are 8.74 and 51.83 percentage points; relative percentage improvements are undefined because both baselines are zero.

\begin{figure}[htbp]\centering
\includegraphics[width=\linewidth]{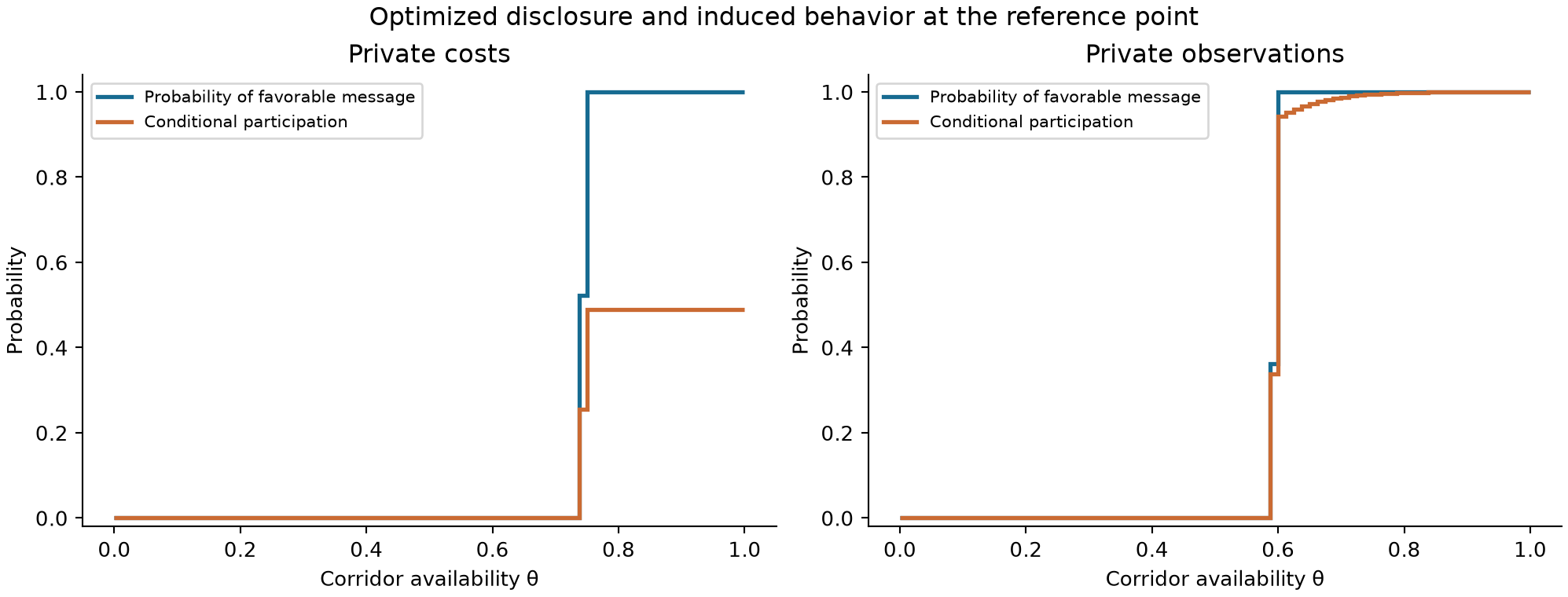}
\caption{Recovered reference signal rules and resulting state-conditional participation. State cells include the additional model boundaries described in the text.}\label{fig:design-signals}
\end{figure}

The private-cost policy sends its productive message with probability 17.83\%, induces posterior mean 0.8066, and supports joining probability 48.99\% after that message. The other message induces no joining. Thus disclosure combines a sufficient expected corridor opportunity with the participation of the lower-cost vehicles. In the observation regime, the productive message occurs with probability 52.87\% and supports the private-observation threshold $t=0.3694$; vehicles still condition their actions on their own observations. The other message supports staying out. The policy itself is a state-dependent probability of each message, rather than a direct command that bypasses private incentives (Figure~\ref{fig:design-signals}).

\begin{figure}[htbp]\centering
\includegraphics[width=\linewidth]{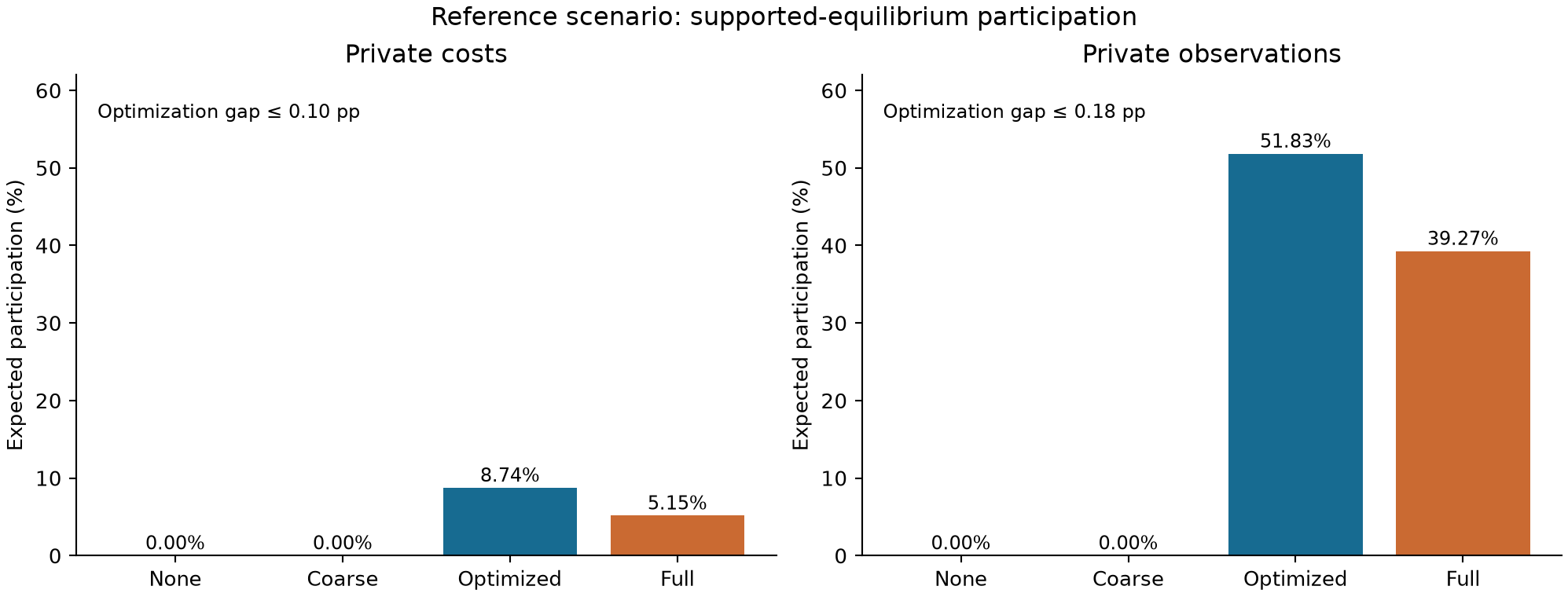}
\caption{Reference participation under the four policies. The stated optimization gaps apply to the restricted numerical design class, not to unrestricted continuous-state design.}\label{fig:design-reference}
\end{figure}

\paragraph{Research question 2: how does private heterogeneity matter?}
Cost dispersion has competing effects. At mean cost \$1.50, moving the coefficient of variation from 0.15 to 0.35 to 0.55 gives optimized participation of 20.50, 8.74, 19.49\%, respectively; all corresponding no-disclosure values are zero. Greater dispersion introduces cheaper potential participants but also more expensive reluctant vehicles, so disclosure gains need not vary monotonically with dispersion. At mean cost \$0.75 the low- and middle-dispersion baselines already have full participation, leaving no possible improvement. At \$3.00 the tested private-cost configurations support no positive participation under the optimized policies.

\begin{figure}[htbp]\centering
\includegraphics[width=\linewidth]{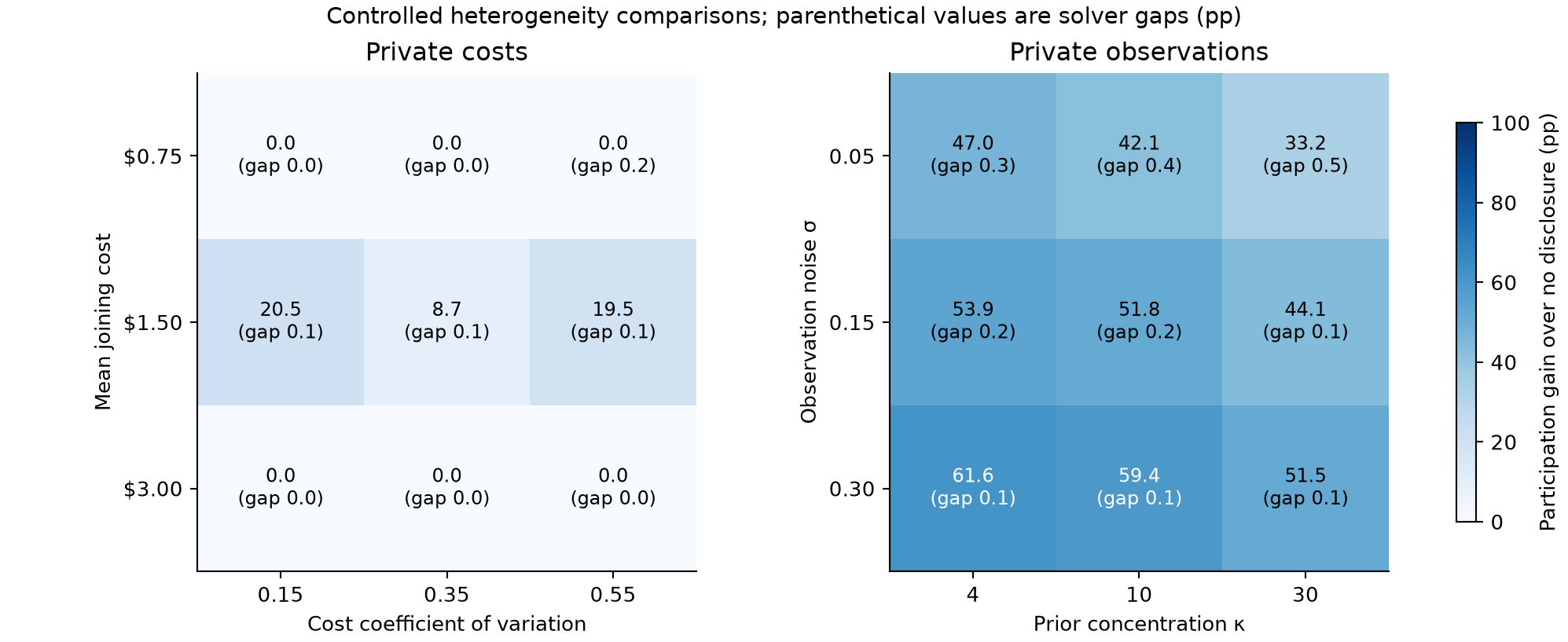}
\caption{Participation gains in the controlled cost and observation treatments. Parenthetical values report each incumbent's remaining numerical upper-bound gap in percentage points.}\label{fig:design-heterogeneity}
\end{figure}

For private observations at prior concentration $\kappa_\theta=10$, increasing noise from $\sigma=0.05$ to 0.15 to 0.30 gives optimized participation of 42.06, 51.83, 59.43\%, respectively, again from zero baselines. Within this calibration, less precise private observations allow the operator's public message to sustain participation across a broader set of opportunities. This is an outcome of the joint information and coordination structure, not a claim that less accurate information universally improves outcomes. Varying prior concentration separately in Figure~\ref{fig:design-heterogeneity} shows that the opportunity distribution also matters. Cross-regime levels are descriptive comparisons at a common mean cost and physical calibration; the two regimes do not have matched preference dispersion or matched private informational content.

\paragraph{Research question 3: when are gains substantial?}
Table~\ref{tab:design-operating-grid} reports every cell in the predefined operating grid. The largest cost-regime gain is 29.01 percentage points and the largest observation-regime gain is 52.74 points. The median gains over the nine equally weighted constructed scenarios are 0.00 and 0.12 points, respectively. These medians describe the experimental grid, not a population average. Returned policies for the short 50-km corridors have zero participation; the remaining observation-design bounds allow at most 0.08 percentage point above these incumbents. At intermediate opportunities disclosure can unlock positive supported participation, while sufficiently favorable long corridors already have near-full baseline participation. Thus large improvements are concentrated in particular operating conditions rather than being representative of every cell. This supports prioritizing a disclosure evaluation where baseline uptake is low but a viable coordinated opportunity exists; it does not identify a universal corridor-length or availability cutoff.

\begin{table}[htbp]\centering\small\setlength{\tabcolsep}{4pt}
\caption{All nine operating scenarios per regime. Participation and gaps are expressed in percentage points or percent as appropriate: $J_0$ and $J_\pi$ columns show $100J$; Gap is the numerical upper bound minus incumbent, multiplied by 100.}\label{tab:design-operating-grid}
\begin{tabular}{rrrrrrrr}
\hline
 & & \multicolumn{3}{c}{Private costs} & \multicolumn{3}{c}{Private observations}\\
$\overline L$ & $\mu_\theta$ & $J_0$ & $J_\pi$ & Gap & $J_0$ & $J_\pi$ & Gap\\\hline
50 & 0.3 & 0.00 & 0.00 & 0.00 & 0.00 & 0.00 & 0.00\\
50 & 0.6 & 0.00 & 0.00 & 0.00 & 0.00 & 0.00 & 0.04\\
50 & 0.9 & 0.00 & 0.00 & 0.00 & 0.00 & 0.00 & 0.08\\
100 & 0.3 & 0.00 & 0.03 & 0.05 & 0.00 & 1.78 & 0.10\\
100 & 0.6 & 0.00 & 8.74 & 0.10 & 0.00 & 51.83 & 0.18\\
100 & 0.9 & 75.75 & 75.78 & 0.10 & 99.68 & 99.80 & 0.10\\
200 & 0.3 & 0.00 & 29.01 & 0.10 & 0.00 & 52.74 & 0.31\\
200 & 0.6 & 100.00 & 100.00 & 0.00 & 99.52 & 99.71 & 0.10\\
200 & 0.9 & 100.00 & 100.00 & 0.00 & 100.00 & 100.00 & 0.00\\
\hline
\end{tabular}
\end{table}

\begin{figure}[htbp]\centering
\includegraphics[width=\linewidth]{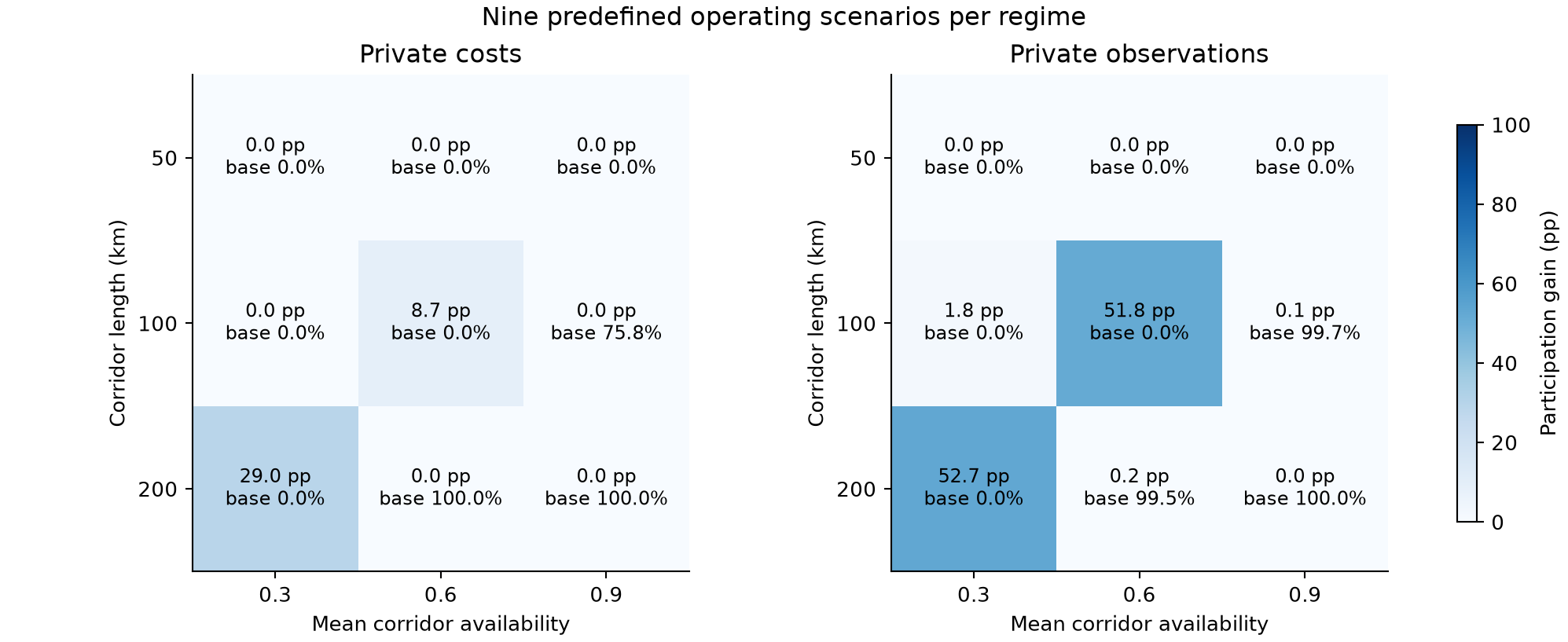}
\caption{Participation improvements across the operating grid, with baseline levels displayed. Gains are conditional on favorable equilibrium selection.}\label{fig:design-operating-grid}
\end{figure}

\paragraph{Transportation benefits and implementation limits.}
At the reference point, the optimized cost policy yields 11.99\% formation probability, 0.118 L gross fuel savings per eligible vehicle, and \$0.039 expected monetary utility per eligible vehicle. The corresponding observation-regime values are 52.86\%, 0.776 L, and \$0.076. In the observation regime, optimized participation exceeds full disclosure and increases gross fuel savings, yet monetary utility falls from \$0.087 under full disclosure to \$0.076. Participation maximization is therefore substantively different from welfare maximization, even in this small example. Access-fuel components are not separately identified by the aggregate cost distribution, so these are gross fuel savings and not net-fuel or emissions estimates.

The additional cohort-size runs are archived with all policy results. At $N=2$, optimized participation is 0.00\% for costs and 21.57\% for observations; at $N=8$, it is 41.42\% and 63.99\%. The latter extrapolates the benefit calibration and is a model sensitivity, not empirical evidence for eight-truck platoons. In every scenario the adverse-selection outcome remains zero under all four policies. Consequently, the numerical gains establish what disclosure can support with favorable coordination, not what a public message guarantees. The next modeling priority is a justified formation or equilibrium-selection mechanism, rather than interpreting these gains as assured adoption.

\paragraph{Reproducible artifacts and remaining extensions.}
The implementation and configuration are stored in \texttt{numerical/}; \texttt{design\_summary.csv} reports all cases, case-level JSON files archive signal probabilities, message statistics, thresholds, numerical bounds, runtime, and paired simulation intervals, and \texttt{design\_audit.json} records verification diagnostics. The supplementary implementation in Section~\ref{sec:numerical-validation} adds message-budget and grid refinements, additive and physical payoff comparisons, constructed-trajectory net-fuel accounting, and off-design incentive checks. Field validation and formal interval-arithmetic certification remain open. 
% END COMPUTED DESIGN RESULTS

% BEGIN SUPPLEMENTARY VALIDATION RESULTS
\subsection{Numerical validation and robustness}\label{sec:numerical-validation}
The supplementary experiments test computational sensitivity, the value of additional messages, and the physical interpretation of the disclosure gains. They retain the main operating assumptions unless a specified factor is changed. The archived configuration fixes the parameter grid and computational budgets; matched-grid two-message controls prevent attributing a state-resolution improvement to a larger message alphabet. These runs supplement the 39 main policy comparisons.

\paragraph{Analytical benchmarks and small-instance verification.}
For the additive example $\theta\sim U[0,1]$, $b=0.2$, and $C_i\sim U[0,1.2]$, the numerical no-disclosure outcome and best incumbent both equal $0.5$, agreeing with Proposition~\ref{prop:no-disclosure}. Its unrestricted analytical optimum follows from that proposition, rather than from the looser numerical allocation bound of $0.507813$. With $C_i\sim U[0.6,1.2]$, the optimizer finds $0.199994$ participation versus zero without disclosure, reproducing the feasible $20\%$ example to within $0.001$ percentage point; the numerical upper bound is $0.200972$. This does not prove the example policy globally optimal in the continuous problem. Exhaustive deviation checks of all 16 homogeneous four-vehicle action profiles agree with the physical equilibrium conditions at $\theta=0.2,0.6,0.9$. On a four-cell physical instance, exhaustive enumeration of 6,561 policies on a nine-point signal-probability grid yields $J=0.059241$, below the continuous-allocation incumbent $0.059542$ and its upper bound $0.060529$. This is an independent small-instance check, not exhaustive verification of the large cases.

\paragraph{State resolution and integration accuracy.}
Table~\ref{tab:supplement-grid} reports the 40-, 80-, and 160-base-cell reference solves, including all additional structural splits. Here $J_C$ and $J_X$ denote participation under private costs and private observations. From 80 to 160 base cells, participation changes by only 0.026 and 0.104 percentage point, respectively. From 40 to 160 cells, the corresponding changes are 0.082 and 0.266 points. These changes are small relative to the reference disclosure gains, although the observation search remains resource-limited and cell refinement alone is not a continuous-state error certificate. Reintegrating all archived supplementary policies with 64 nodes per cell changes participation by less than $10^{-7}$. Conditional indifference residuals and checked observation-incentive violations are below $10^{-6}$ in payoff units. Independent 80,000-round simulations of the supplementary optimized policies agree with integrated participation within two standard errors for the saved seed. The audit also verifies realized platooning fractions and vehicle-distance, and keeps sampling errors separate from optimization bounds.
\begin{table}[htbp]\centering\small
\caption{Reference state-grid refinement. Both regimes use the same structural cell splits; gaps concern each cellwise two-message problem.}\label{tab:supplement-grid}
\begin{tabular}{lrrrrr}
\hline
Base cells & Actual cells & $J_C$ (\%) & Gap (pp) & $J_X$ (\%) & Gap (pp)\\\hline
40 & 42 & 8.682 & 0.099 & 51.671 & 0.154\\
80 & 82 & 8.737 & 0.100 & 51.833 & 0.179\\
160 & 162 & 8.763 & 0.094 & 51.937 & 0.169\\
\hline
\end{tabular}
\end{table}

\paragraph{A richer public message alphabet helps more in the observation examples.}
For a four-message budget, private-cost design uses a 3,200-interval posterior-mean LP relaxation as an upper bound, allowing unrestricted message count in that relaxation. Feasible policies are recovered by merging allocations and recomputing greatest cost fixed points. Private-observation design first selects at most four columns from a threshold library by MILP, then refines their thresholds through fixed-threshold LPs. Continuous-threshold interval relaxations, including necessary incentive tests outside each threshold interval, give upper bounds. Constant-action messages are retained. A library MILP uses a 30-second search budget; its finite-library optimality bound is not substituted for the continuous-threshold bound.

Table~\ref{tab:supplement-messages} compares the budgets on matched state grids. The four-message cost incumbents retain two active messages and essentially the same participation, with remaining upper-bound gaps below 0.1 percentage point. This bounds the potential extra gain in these examples, without proving that two messages generally suffice. In the observation regime, four active messages raise reference participation from $51.83\%$ to $53.12\%$. At 200 km and prior mean 0.3, participation rises from $52.93\%$ to $54.61\%$. The four-message observation upper-bound gaps are 0.89 and 1.10 percentage points. These are feasible improvements with explicitly unfinished optimality gaps; the richer-alphabet policies are not claimed to attain the unrestricted optimum. The marginal gain is much smaller than the gain from replacing no public disclosure with the original two-message design.
Both four-message observation incumbents exceed the corresponding two-message upper bounds on their matched grids. Thus their improvements cannot be explained solely by an unfinished two-message search within the stated numerical class.
\begin{table}[htbp]\centering\small
\caption{Message-budget sensitivity on matched state grids. Participation columns are percentages; four-message incumbents may use fewer messages. Bounds relax continuous thresholds or posterior means and are not proofs that two messages suffice.}\label{tab:supplement-messages}
\begin{tabular}{llrrrr}
\hline
Scenario & Private type & Two msg. & Four msg. & Used & Gap (pp)\\\hline
Reference & Costs & 8.74 & 8.74 & 2 & 0.04\\
Reference & Observations & 51.83 & 53.12 & 4 & 0.89\\
200 km, $\mu_\theta=0.3$ & Costs & 29.02 & 29.02 & 2 & 0.08\\
200 km, $\mu_\theta=0.3$ & Observations & 52.93 & 54.61 & 4 & 1.10\\
\hline
\end{tabular}
\end{table}

\begin{figure}[htbp]\centering
\includegraphics[width=\linewidth]{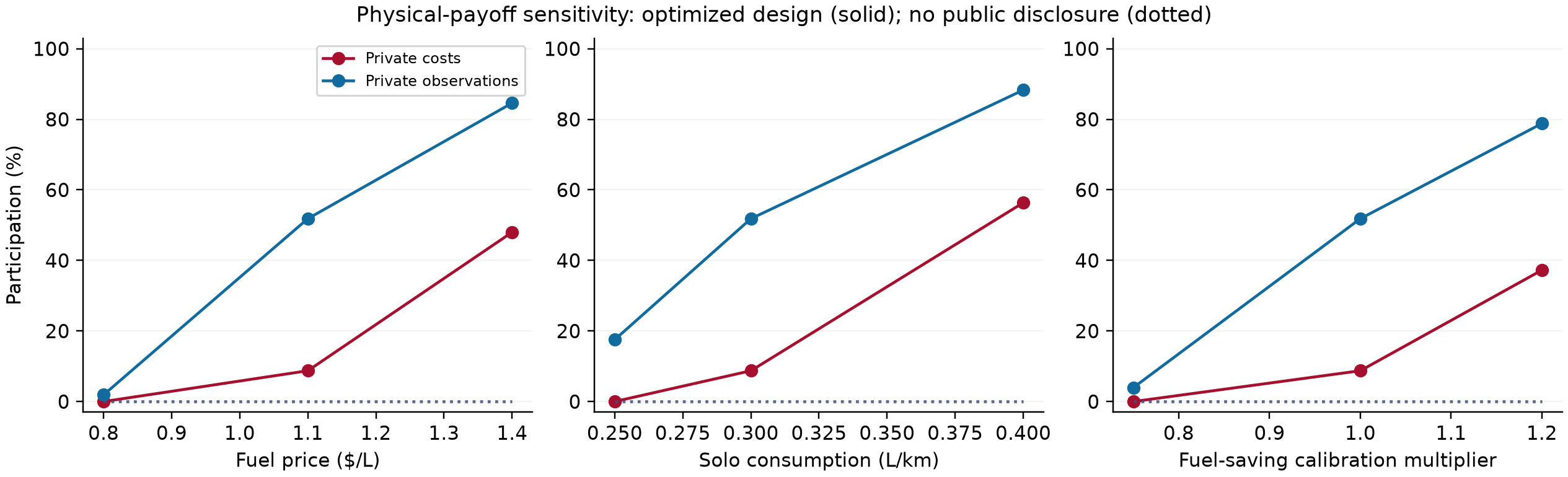}
\caption{One-factor physical calibration sensitivities. Solid curves show two-message incumbents and dotted curves show no public disclosure. These are controlled scenario responses, not estimated fuel-price or adoption elasticities.}\label{fig:supplement-sensitivity}
\end{figure}

\paragraph{Fuel value and joining incentives govern where disclosure can help.}
Figure~\ref{fig:supplement-sensitivity} varies one physical calibration factor at a time, holding the prior, cohort, private heterogeneity, and mean joining cost fixed. Raising fuel price from \$0.80 to \$1.40/L increases optimized cost-regime participation from zero to $47.99\%$ and observation-regime participation from $1.90\%$ to $84.60\%$. Varying the leader/follower targets from $3\%/6\%$ to $4.8\%/9.6\%$ changes these outcomes from zero to $37.26\%$, and from $3.96\%$ to $78.81\%$, respectively. Solo-consumption scenarios of 0.25--0.40 L/km span $0.012\%$--$56.36\%$ for costs and $17.55\%$--$88.29\%$ for observations. No-disclosure participation remains zero in these sensitivity cases, so levels also measure absolute disclosure gains. Their largest remaining design gap is below 0.70 percentage point. These results reinforce the operating-grid finding: information can support coordination when expected benefits can cover joining costs, but does not substitute for an economically viable opportunity. The constant-speed model identifies the product of total-fuel saving and fuel value; changing the assumed aerodynamic share while recalibrating drag reductions to the same total-fuel targets leaves that product unchanged.

\paragraph{Local additive approximations can give misleading disclosure policies.}
Table~\ref{tab:supplement-additive} uses the fixed interpolation expansion described in Section~\ref{sec:numerical-types}. At 100 km, the additive-designed policies predict $34.45\%$ participation with private costs and $59.08\%$ with private observations. Holding each public signal rule fixed and recomputing physical equilibria yields zero in both cases. At 200 km, by contrast, the physical responses remain $100\%$ and $99.52\%$. The approximation therefore performs differently across operating regions. This comparison includes integer-size interpolation and local linearization; it is not an error estimate for a Taylor expansion of the exact polynomial $\mathcal R_N$. Its analytical tractability is useful, but its nonzero isolated-joiner benefit can alter the very formation incentives on which persuasion depends. A signal should be evaluated with the exact physical payoff when outcomes may cross that formation boundary; tightening its additive-model optimization gap alone does not correct this error.
\begin{table}[htbp]\centering\small
\caption{Additive-designed signals evaluated with their own payoff and with recomputed physical-payoff equilibria. Participation is percent. The gap bounds additive design only; it does not bound approximation error.}\label{tab:supplement-additive}
\begin{tabular}{llrrr}
\hline
Corridor (km) & Private type & Predicted $J$ & Physical $J$ & Gap (pp)\\\hline
50 & Costs & 2.45 & 0.00 & 0.10\\
50 & Observations & 0.00 & 0.00 & 0.00\\
100 & Costs & 34.45 & 0.00 & 0.12\\
100 & Observations & 59.08 & 0.00 & 0.98\\
200 & Costs & 100.00 & 100.00 & 0.00\\
200 & Observations & 99.97 & 99.52 & 0.03\\
\hline
\end{tabular}
\end{table}

\paragraph{Feasible access costs and net fuel accounting.}
The trajectory check uses a 29,484-kg vehicle, air density $1.225$ kg/m$^3$, drag area $6$ m$^2$, rolling coefficient $0.006$, and fuel energy $35.8$ MJ/L. Efficiency and idling fuel rate are calibrated to the reference constant-speed consumption and aerodynamic share in~\eqref{eq:traction-model}. These additional engineering inputs are scenario assumptions. The 501 access paths sample a 50--250 m catch-up distance, 60--240 s waiting time, and a 10--20\% speed increment above 55 mph, subject to $0.30$ m/s$^2$ acceleration and a 70-mph speed ceiling. Each path starts and ends at the reference speed. Excess fuel is measured against solo travel over the same access distance; additional time deducts the corresponding solo travel time. All generated paths satisfy the specified speed, acceleration, delay, and closure checks.

A continuous piecewise-linear CDF passes through the minimum, 10th, 25th, 50th, 75th, and 90th percentiles, and maximum of the sampled costs. Its mean is \$1.718 and CV is 0.334. Fuel and delay components are interpolated at the same knots, preserving the identity $\mathcal C=p_f F_{\mathrm{access}}+wT_{\mathrm{delay}}$; integrating these components only over joining types avoids counting access costs twice. A matched-mean, matched-variance uniform cost distribution gives $0.349\%$ optimized participation, compared with $0.275\%$ for the access CDF, with design gaps of 0.043 and 0.089 percentage point. The bounds overlap, so this small difference does not establish a reliable ranking of CDF shapes. More consequentially, applying the original \$1.50 reference signals to the higher access-cost environment yields zero supported participation in both regimes. Redesigning at the new cost distribution, or at its mean in the common-cost observation regime, restores $0.275\%$ and $25.88\%$ at constant speed. Thus cost recalibration matters even when the mean changes by only about $15\%$.

For downstream validation, all policies face the same flat, mild, or hilly spatial speed/grade profile. Mild and hilly profiles vary speed by $\pm5\%$ and $\pm10\%$ about 55 mph and grade by $\pm0.5\%$ and $\pm1\%$, using 20- and 25-km wavelengths. The first $\theta\overline L$ of the common corridor is available for platooning; the remainder is solo. Fuel follows the positive-traction model, averaging actual leader and follower fuel use. The flat profile reproduces the constant-speed benefit to numerical precision, and halving the spatial step from 100 to 50 m changes the monetary benefit by less than \$0.001. With reference costs held fixed, varying the profile changes participation from $8.74\%$ to $9.55\%$ for costs and $51.83\%$ to $52.72\%$ for observations. This effect is much smaller than that of replacing the joining-cost calibration.

\begin{figure}[htbp]\centering
\includegraphics[width=\linewidth]{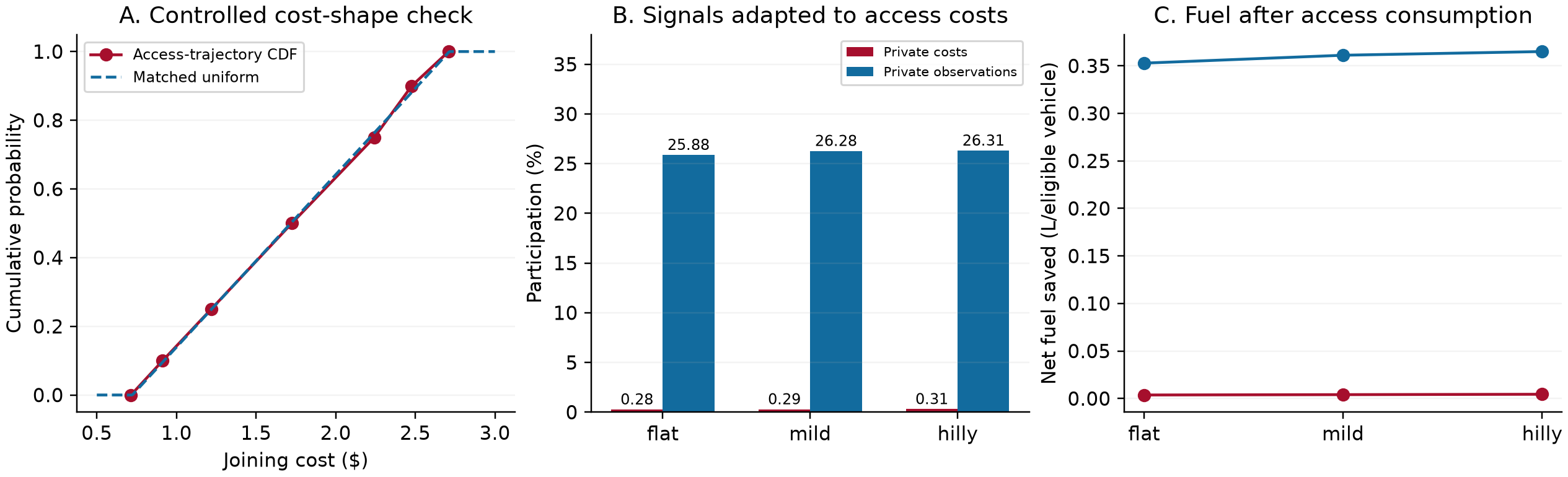}
\caption{Constructed-trajectory validation: matched-moment cost distributions, recomputed participation under access-adapted signals, and fuel saved after access consumption. All-out remains an equilibrium.}\label{fig:supplement-trajectories}
\end{figure}

Table~\ref{tab:supplement-traffic} and Figure~\ref{fig:supplement-trajectories} evaluate the access-adapted signals on these profiles. Signals are designed using the constant-speed payoff and then held fixed, with equilibria recomputed for each profile; these are not trajectory-specific optimal policies. Net savings subtract the fuel used for access: the observation outcomes save 0.353--0.365 L per eligible vehicle while incurring about 37--38 s additional time per eligible vehicle. The low cost-regime participation yields only 0.0037--0.0045 L net savings. These constructed paths make fuel/time accounting concrete but do not establish network-level traffic or emissions effects.
\begin{table}[htbp]\centering\small
\caption{Recomputed physical outcomes under signals adapted to trajectory-based access costs. All fuel, delay and monetary values are per eligible vehicle. Profiles and access paths are constructed scenarios, not field data.}\label{tab:supplement-traffic}
\begin{tabular}{llrrrr}
\hline
Profile & Private type & $J$ (\%) & Net fuel (L) & Delay (s) & Utility (\$)\\\hline
Flat & Costs & 0.275 & 0.0037 & 0.27 & 0.0014\\
Flat & Observations & 25.880 & 0.3528 & 37.06 & 0.0207\\
Mild & Costs & 0.291 & 0.0041 & 0.29 & 0.0016\\
Mild & Observations & 26.278 & 0.3612 & 37.63 & 0.0241\\
Hilly & Costs & 0.308 & 0.0045 & 0.32 & 0.0018\\
Hilly & Observations & 26.313 & 0.3650 & 37.68 & 0.0279\\
\hline
\end{tabular}
\end{table}

\paragraph{Persuasiveness must be checked again when conditions change.}
Holding the reference signal rule fixed, reducing prior concentration from 10 to 4 changes greatest supported participation to $19.48\%$ with costs and $53.29\%$ with observations; raising concentration to 30 yields zero in both. Reducing the mean cost by $10\%$ gives $13.38\%$ and $52.87\%$, while increasing it by $10\%$ again yields zero. Recomputed equilibria respect the perturbed incentives, but the originally prescribed cutoffs can fail them: maximum checked profitable deviations range from \$0.017 to \$0.149 across these perturbations. A previously persuasive recommendation is therefore not automatically persuasive after its prior or cost calibration changes. These tests concern changed primitives known to vehicles; they are not a solution to a distributionally robust design problem.

\paragraph{Realized risk and reproducibility.}
At the reference two-message policy, independent round simulations estimate negative realized net payoffs for about $29.7\%$ of joining vehicles in the cost regime and $25.5\%$ in the observation regime, despite interim optimality. Integrated realized-platooning fractions are $7.58\%$ and $51.83\%$, with 24.45 and 148.25 platooning vehicle-km per four-vehicle decision round. These diagnostics distinguish ex ante persuasion, viable formation, and realized benefit. The supplementary scripts, frozen configuration, case-level policies, CSV summary, independent audit, and source/result hashes are stored in \texttt{numerical/}. Numerical bounds remain floating-point and restricted to the stated state cells, message budgets, and threshold range. Every physical experiment still admits everyone staying out, so neither these refinements nor the trajectory checks remove the implementation limit in Corollary~\ref{cor:physical-implementation}.
% END SUPPLEMENTARY VALIDATION RESULTS

\FloatBarrier
\section{Discussion and Conclusion}\label{sec:discussion}
\subsection{Implementation and meaning of behavioral stability}
A public signal could disclose a category of platooning opportunity based on operator observations. A vehicle then combines this message with its own joining cost or state observation. Persuasiveness relies on a known committed disclosure rule and correct Bayesian interpretation. The modeled stability concerns interim participation incentives. It does not establish stable position assignments, equal realized benefits, or compliance with a subsequent control protocol.

\subsection{Policy implications and their evidential basis}\label{sec:policy-implications}
\paragraph{Match disclosure design to the information environment.}
Proposition~\ref{prop:information-structure} gives the analytical basis for treating private operating costs and private corridor assessments differently. In the independent-cost specification, the operator can compute the symmetric response correspondence using each message's posterior mean and the cost distribution; a consistent selection rule remains necessary when several roots exist. In the observation specification, a mean-only representation can miss equilibria and participant-count dependence. The platform should therefore estimate which source of private heterogeneity is relevant and evaluate state-dependent formation probabilities. The noise sensitivity in Section~\ref{sec:numerical-design-results} motivates assessing public disclosure jointly with existing private evidence; it is not a recommendation to reduce observation accuracy.

\paragraph{Target viable opportunities and disclose the objective.}
Table~\ref{tab:design-reference} shows reference gains over full disclosure of 3.59 and 12.56 percentage points in the cost and observation regimes, respectively, under the common favorable-selection convention. Table~\ref{tab:design-operating-grid} locates the largest baseline improvements at intermediate opportunities, while its median improvements are only 0.00 and 0.12 points. An operator can use this pattern to screen where a detailed disclosure optimization is worth evaluating, rather than assume a broadly positive effect. These are conditional model findings, not estimated deployment effects. If the platform values uptake, it should report commitments alongside formation and gross fuel savings; if it values welfare, the observed utility loss relative to full disclosure motivates a different objective or additional welfare constraints. The present computations demonstrate the tradeoff, but do not solve that alternative design problem or quantify net emissions. The constructed-trajectory checks in Section~\ref{sec:numerical-validation} further show why access fuel must be deducted and why a signal should be recalibrated when joining costs change; their net-fuel estimates concern those scenarios only.

\paragraph{Evaluate formation commitments before promising adoption.}
Corollary~\ref{cor:physical-implementation} establishes the policy boundary independently of the numerical calibration: strictly positive joining costs and the absence of a guaranteed partner preserve an all-out equilibrium under every public signal. A public message alone therefore cannot guarantee a positive worst-equilibrium participation rate. A minimum-participant commitment, an initial partner, or a complementary matching or transfer mechanism is a possible extension; each changes the incentives and must be shown to be feasible and to eliminate the relevant low-participation outcome. This paper proves the need for an additional implementation basis when guaranteed formation is desired, not the sufficiency of any particular added mechanism.

\subsection{Limitations and extensions}
The participation objective is not a welfare objective: more participation need not imply more fuel saving or larger total utility. The physical specialization fixes cruising speed, gap, a rendezvous protocol, and exchangeable position assignment. The effective-distance state and empirical drag and fuel parameters still require calibration. The additive approximation has local validity and does not preserve the physical model's formation boundary. Independent costs and conditionally independent observation noises simplify inference; correlated types, cost-state dependence, or both forms of private heterogeneity would require additional analysis. Network compatibility among vehicles, endogenous state variables, repeated interactions, and post-formation leader allocation are further extensions. A continuum approximation should be justified against finite-vehicle outcomes when used.

\subsection{Conclusions}
Public information is a complementary instrument for voluntary platoon formation. Its value depends on how vehicles combine public messages with their own joining costs or corridor assessments. The two regimes require different equilibrium representations and disclosure computations, and they generate different formation risks even at similar average participation.

The experiments show that optimized disclosure can outperform both silence and full revelation, particularly when coordinated formation is viable but baseline uptake is low. They also show why deployment requires care: private-cost calibration, physical formation incentives, and the operator's objective affect the apparent gains. Public messages support participation under the stated equilibrium convention; they do not guarantee that vehicles coordinate on that outcome. The policy priority is to combine disclosure with a credible formation process and to evaluate fuel, delay, and welfare alongside commitments.

\bibliographystyle{plainnat}
\bibliography{reference}
\end{document}